\documentclass[11pt,a4paper]{article}

\pdfoutput = 1

\usepackage{jcappub}
\usepackage{multirow}
\pdfoutput=1
\usepackage[utf8]{inputenc}
\usepackage{float}
\usepackage{hyperref}
\usepackage{amsmath}
\usepackage{graphicx}
\usepackage{dcolumn}
\usepackage{bm}
\usepackage{epsfig}
\usepackage{amssymb,latexsym,mathrsfs}
\usepackage{graphicx}
\usepackage{color}
\usepackage{booktabs}

\usepackage{comment}

\definecolor{myred}{RGB}{180,50,28}
\definecolor{myblue}{RGB}{2,50,180}
\definecolor{mygreen}{RGB}{2,150,80}

\hypersetup{colorlinks=true,
            citecolor=blue,
            linkcolor=blue,
            urlcolor=blue}

\newcommand{\be}{\begin{equation}}
\newcommand{\ee}{\end{equation}}
\newcommand{\bea}{\begin{eqnarray}}
\newcommand{\eea}{\end{eqnarray}}

\title{Biased tracers, Hybrid Effective Field Theory and Modified Gravity}

\author[a,1]{Guilherme Brando,\note{Corresponding author.}}
\author[b]{Baojiu Li,}
\author[c,d,e]{Kazuya Koyama}

\affiliation[a]{CBPF - Brazilian Center for Research in Physics, \\
Xavier Sigaud st. 150, zip 22290-180, Rio de Janeiro, RJ, Brazil}
\affiliation[b]{Institute for Computational Cosmology, Department of Physics,\\
Durham University, Durham DH1
3LE, UK}
\affiliation[c]{Institute of Cosmology and Gravitation, University of Portsmouth,\\ Dennis Sciama Building, Burnaby Road, Portsmouth PO1 3FX, United Kingdom}
\affiliation[d]{Kavli IPMU (WPI), UTIAS, The University of Tokyo, Kashiwa, Chiba 277-8583, Japan}
\affiliation[e]{Yukawa Institute for Theoretical Physics, Kyoto University, Kyoto 606-8502, Japan}

\emailAdd{gbrando@cbpf.br}

\abstract{The modelling of the power spectrum of biased tracers has become a central topic in the analysis of modern cosmological galaxy surveys. Perturbative templates formulated in both Eulerian and Lagrangian frameworks have been extensively developed over the last decades, with their implementation in $\Lambda$CDM thoroughly investigated and validated. In parallel, approaches combining perturbation theory with the output of dark-matter-only simulations have emerged as powerful tools for modelling the nonlinear regime, most notably the Hybrid Effective Field Theory (HEFT) framework~\cite{Modi:2019qbt}. In this work, we discuss the perturbative biased expansion within the local Lagrangian bias scheme and its implementation in the HEFT framework for modified gravity cosmologies. We focus on $f(R)$ gravity, a theory characterized by scale-dependent growth and chameleon screening, making it one of the most challenging scenarios for the computation of Lagrangian Perturbation Theory growth functions and for the generation of accurate numerical simulations. We present a detailed overview of the ingredients required to compute loop-corrected biased power spectra analytically and compare these predictions against fully non-perturbative simulation results. Finally, we propose a strategy to extend existing HEFT-based $\Lambda$CDM emulators, such as \texttt{bacco} and \texttt{Aemulus}, to beyond-$\Lambda$CDM cosmologies.}
\begin{document}
\maketitle
\flushbottom

\section{Introduction} \label{intro}

The era of Stage-IV galaxy surveys is now upon us. The Dark Energy Spectroscopic Instrument (DESI)~\cite{DESI:2016fyo}, the Vera Rubin Legacy Survey of Space and Time~\cite{LSSTScience:2009jmu} (LSST), the Nancy Grace Roman Space Telescope~\cite{roman1,roman2} (Roman), and the Euclid mission~\cite{EUCLID:2011zbd} will provide unprecedented data probing both the geometry and the growth of structure in our Universe. From the perspective of structure formation, the nonlinear matter power spectrum, relevant for cosmic shear and CMB lensing analyses, together with the biased tracer power spectrum, are among the fundamental observables that must be modelled theoretically. As observational precision continues to improve, advancing the theoretical description of these observables in both $\Lambda$CDM and beyond-$\Lambda$CDM cosmologies becomes increasingly important.

One of the key challenges is the computation of the power spectrum of biased tracers, such as galaxies. Since galaxies trace the underlying dark matter distribution in a nontrivial way, accurate and efficient modelling of the galaxy power spectrum is essential for analyses of galaxy clustering and galaxy--galaxy lensing. As in the case of the matter power spectrum, there are two main avenues to compute these quantities theoretically: perturbative methods~\cite{Bernardeau:2001qr,Desjacques:2016bnm} and cosmological simulations~\cite{Wechsler:2018pic}. The former offers greater flexibility for exploring different cosmological models and are generally much less computationally expensive than running large suites of simulations. The latter, while more computationally demanding, also require a nontrivial connection between the simulation output -- namely the positions and velocities of dark matter particles -- and the observed galaxy distribution. Several approaches exist to populate dark-matter-only simulations with galaxies, including Halo Occupation Distribution (HOD)~\cite{Zheng:2004id}, SubHalo Abundance Matching (SHAM)~\cite{Kravtsov:2003sg, Tasitsiomi:2004pt}, and semi-analytic models~\cite{White:1991mr,Kauffmann:1993umz,Cole:2000ex}. In recent years, hydrodynamical simulations~\cite{Crain:2023xap} have also become increasingly common, with GPU-based methods helping to reduce their computational cost.

Broadly speaking, these two approaches represent a trade-off between tractability and predictive power: perturbation theory provides a more efficient framework for parameter inference and MCMC analyses, while simulations, whether dark-matter-only or hydrodynamical, offer a more complete description of nonlinear small-scale physics. Motivated by bridging this gap, Ref.~\cite{Modi:2019qbt} proposed an approach that extends the validity of perturbative templates by combining a biased tracer expansion in Lagrangian Perturbation Theory with the output of dark-matter-only simulations. This method, known as Hybrid Effective Field Theory (HEFT), has since been implemented in several emulators~\cite{Zennaro:2021bwy,DeRose:2018xdj,Zhou:2025iiu} and widely applied in the literature~\cite{LSSTDarkEnergyScience:2023qfp,Kokron:2021faa,deBelsunce:2025gci,Hadzhiyska:2021xbv,Banerjee:2021cmi,Rubiola:2025ntk,Pellejero-Ibanez:2022efv,Garcia-Garcia:2024gzy}.

A common feature of all these works on HEFT and its applications is that they have been developed primarily within $\Lambda$CDM and mild extensions of it, such as $w_{0}w_{a}$CDM. In this work, we are interested in extending the theoretical predictions of the HEFT basis spectra to modified theories of gravity. To do so, we revisit the literature on the Lagrangian local biasing scheme, which forms the theoretical foundation of the HEFT approach, in these models. While the majority of perturbation theory works have focused on the development and computation of observable quantities in the Eulerian frame~\cite{Senatore:2014via,Senatore:2017pbn,LEFT,Baldauf:2015xfa,Vlah:2015zda,Blas:2016sfa,Ding:2017gad,Ivanov:2018gjr}, the biased expansion in Lagrangian coordinates and fields was first introduced in~\cite{Matsu2}, extending the loop-corrected matter power spectrum formalism of~\cite{Matsu1}. Building on these works, the Lagrangian approach has since been significantly developed in~\cite{CLEFT,LEFT,Vlah:2014nta,Vlah:2016bcl,Chen:2020fxs,White:2014gfa,Kokron:2021faa,DeRose:2022zfu,Chudaykin:2020aoj,Nishimichi:2020tvu}.

One of the appealing aspects of this framework is its natural extension to generalized cosmologies, such as modified gravity theories. In this context, Refs.~\cite{Aviles:2017aor,AA2,Valogiannis:2019xed} carefully derived the equations of motion for a class of modified gravity theories, allowing the computation of the displacement field up to third order, which is necessary for evaluating the LPT quantities. In the present work, we revisit and extend these results to develop an efficient numerical framework for computing LPT quantities with growth factors that are both scale- and time-dependent, as well as the associated loop integrals required to ultimately predict the biased tracer power spectrum. We compare and validate our implementation against existing codes in the literature, and explicitly quantify the limitations of the Einstein--de Sitter approximation in generalized cosmologies. Once with the theoretical analytical LPT results we can then compute the HEFT results and follow the same strategy as presented in~\cite{Zennaro:2021bwy,DeRose:2018xdj,Zhou:2025iiu}

This paper is structured as follows. In Section~\ref{sec:rev_pt}, we begin with a brief review of perturbation theory in General Relativity (GR). In Section~\ref{sec:lpt}, we review the dynamics of Lagrangian Perturbation Theory and describe how the one-loop corrected matter power spectrum is computed in this framework. We then discuss the modified gravity extension of LPT in Section~\ref{sec:mg}. Next, in Section~\ref{sec:biased}, we review the fundamentals of biased tracers in LPT, and in Section~\ref{sec:heft} the HEFT formalism and present our results in both modified gravity and GR, considering theoretical predictions from CLPT~\cite{CLEFT} as well as HEFT predictions calibrated on \textit{N}-body simulations. Finally, we summarize and discuss our conclusions in Section~\ref{sec:concl}.

%\section{Methodology}\label{sec:method}

%In this section we will present in a schematic way how we constructed the necessary quantities for the implementation of the Hybrid Effective Field Theory in Modified Gravity. 

\section{Perturbation Theory in General Relativity}\label{sec:rev_pt}

We begin by briefly reviewing the formulation of cosmological perturbation theory within the Eulerian framework. In the Eulerian picture, cold dark matter (CDM) dynamics are treated as those of a fluid, where the density contrast $\delta(\boldsymbol{x},t)$ and the velocity field $\boldsymbol{v}(\boldsymbol{x},t)$ are the fundamental variables. The equations of motion are obtained by describing how the matter fluid flows through a fixed three-dimensional comoving coordinate grid, $\boldsymbol{x}$, as a function of time $t$. 

It is well known that, during matter domination, CDM dynamics on scales well inside the cosmological horizon are accurately described by a Newtonian fluid approach. On such scales, relativistic corrections arising from gauge choices and additional relativistic terms are subdominant and typically have negligible impact on the dynamics. In the Eulerian framework, the equations of motion for $\delta(\boldsymbol{x},t)$ and $\boldsymbol{v}(\boldsymbol{x},t)$ are given by the continuity, Euler and Poisson equation:
\begin{align}
\frac{\partial \delta(\boldsymbol{x},\tau)}{\partial \tau}
&+ \nabla \cdot \Big[(1+\delta(\boldsymbol{x},\tau))\,\boldsymbol{v}(\boldsymbol{x},\tau)\Big] = 0 ,
\label{eq:continuity} \\
\frac{\partial \boldsymbol{v}(\boldsymbol{x},\tau)}{\partial \tau}
&+ \mathcal{H}(\tau)\,\boldsymbol{v}(\boldsymbol{x},\tau)
+ \big[\boldsymbol{v}(\boldsymbol{x},\tau)\cdot\nabla\big]\boldsymbol{v}(\boldsymbol{x},\tau)
= -\nabla \Phi(\boldsymbol{x},\tau) ,
\label{eq:euler}\\
\nabla^2 \Phi(\boldsymbol{x},\tau) 
&= 4\pi G\,a^2(\tau)\,\bar{\rho}_m(\tau)\,\delta(\boldsymbol{x},\tau) ,
\label{eq:poisson}
\end{align}
where $\tau$ is the conformal time, defined through $dt = a d\tau$, and $\mathcal{H}$ is the conformal Hubble factor, $\mathcal{H}=aH$. In Fourier space, these equations are written as
\begin{align}
&\frac{\partial \tilde{\delta}(\boldsymbol{k},\tau)}{\partial \tau}
+ \tilde{\theta}(\boldsymbol{k},\tau)
=
- \int d^3\boldsymbol{k}_1 \, d^3\boldsymbol{k}_2 \,
\delta_D(\boldsymbol{k} - \boldsymbol{k}_{12}) \,
\alpha(\boldsymbol{k}_1,\boldsymbol{k}_2)\,
\tilde{\theta}(\boldsymbol{k}_1,\tau)\,
\tilde{\delta}(\boldsymbol{k}_2,\tau) ,
\label{eq:delta_k}\\
&\frac{\partial \tilde{\theta}(\boldsymbol{k},\tau)}{\partial \tau}
+ \mathcal{H}(\tau)\,\tilde{\theta}(\boldsymbol{k},\tau)
+ \frac{3}{2}\,\Omega_m(\tau)\,\mathcal{H}^2(\tau)\,\tilde{\delta}(\boldsymbol{k},\tau)
=\nonumber\\
&- \int d^3\boldsymbol{k}_1 \, d^3\boldsymbol{k}_2 \,
\delta^{\rm D}(\boldsymbol{k} - \boldsymbol{k}_{12})
\beta(\boldsymbol{k}_1,\boldsymbol{k}_2)\,
\tilde{\theta}(\boldsymbol{k}_1,\tau)\,
\tilde{\theta}(\boldsymbol{k}_2,\tau) ,
\label{eq:theta_k}
\end{align}
where we have already written the velocity contribution in terms of the velocity divergence field, $\theta(\boldsymbol{x},\tau)$.\footnote{The velocity field can be decomposed into longitudinal and transverse components. However, the rotational (transverse) component decays as $a^{-1}$, so we retain only the irrotational part, characterized by the velocity divergence $\theta(\boldsymbol{x},\tau)=\boldsymbol{\nabla}\cdot\boldsymbol{v}(\boldsymbol{x},\tau)$.} The functions
\begin{equation}\label{eq:alpha_beta_kernels}
\alpha(\boldsymbol{k}_{1}, \boldsymbol{k}_{2}) =
\frac{(\boldsymbol{k}_{1}+\boldsymbol{k}_{2})\cdot\boldsymbol{k}_{1}}{k_{1}^{2}},
\qquad
\beta(\boldsymbol{k}_{1}, \boldsymbol{k}_{2}) =
\frac{|\boldsymbol{k}_{1}+\boldsymbol{k}_{2}|^{2}
(\boldsymbol{k}_{1}\cdot\boldsymbol{k}_{2})}
{2k_{1}^{2}k_{2}^{2}},
\end{equation}
are the mode-coupling kernels.

Perturbation theory consists of expanding the full solutions as
\begin{align}
\delta(\boldsymbol{k},\tau)
&=
\sum_{n=1}^{\infty}
\delta^{(n)}(\boldsymbol{k},\tau),
\qquad
\theta(\boldsymbol{k},\tau)
=
\sum_{n=1}^{\infty}
\theta^{(n)}(\boldsymbol{k},\tau),
\end{align}
which are then inserted into Eqs.~(\ref{eq:delta_k})--(\ref{eq:theta_k}) to obtain
\begin{align}\label{eq:delta_theta_n}
\delta^{(n)}(\boldsymbol{k}, \tau)
&=
\int_{\boldsymbol{k}_{1} \dots \boldsymbol{k}_{n}}
\delta^{\rm D}
(\boldsymbol{k}_{1}+ \dots + \boldsymbol{k}_{n} - \boldsymbol{k})
F_{n}(\boldsymbol{k}_{1}, \dots, \boldsymbol{k}_{n};\tau)
\left[
\delta_{\rm lin}(\boldsymbol{k}_{1},\tau)
\dots
\delta_{\rm lin}(\boldsymbol{k}_{n},\tau)
\right],
\\
\theta^{(n)}(\boldsymbol{k}, \tau)
&=
\int_{\boldsymbol{k}_{1} \dots \boldsymbol{k}_{n}}
\delta^{\rm D}
(\boldsymbol{k}_{1}+ \dots + \boldsymbol{k}_{n} - \boldsymbol{k})
G_{n}(\boldsymbol{k}_{1}, \dots, \boldsymbol{k}_{n};t)
\left[
\delta_{\rm lin}(\boldsymbol{k}_{1},\tau)
\dots
\delta_{\rm lin}(\boldsymbol{k}_{n},\tau)
\right].
\end{align}
Here,
\begin{equation}
\int_{\boldsymbol{k}_{1}\dots \boldsymbol{k}_{n}}
=
\frac{1}{(2\pi)^{3(n-1)}}
\int d^{3}\boldsymbol{k}_{1}
\dots
d^{3}\boldsymbol{k}_{n},
\end{equation}
and $F_n(\boldsymbol{k}_{1}, \dots, \boldsymbol{k}_{n};\tau)$ and $G_n(\boldsymbol{k}_{1}, \dots, \boldsymbol{k}_{n};\tau)$ are the perturbation theory kernels that determine the coupling between Fourier modes while respecting the symmetries of the theory. Written in this form, the kernels depend both on scale and time. Their explicit evolution can be obtained recursively, order by order, by substituting the perturbative expansions into Eqs.~(\ref{eq:delta_k})--(\ref{eq:theta_k}) and deriving the corresponding equations of motion for $F_n$ and $G_n$.

In $\Lambda$CDM and in some of its extensions where the growth function is scale independent, one can assume that the time and scale dependence of the perturbation theory kernels approximately decouple. In this case, we may write
\begin{align}
    \delta(\boldsymbol{k}, \tau)
    &=
    \sum_{n=1}^{\infty}
    D_{1}^{n}(\tau)\,\delta^{(n)}(\boldsymbol{k}),
    \qquad
    \theta(\boldsymbol{k}, \tau)
    =
    -\mathcal{H}f
    \sum_{n=1}^{\infty}
    D_{1}^{n}(\tau)\,\theta^{(n)}(\boldsymbol{k}),
\end{align}
where $D_{1}(\tau)$ is the linear growth factor and $f$ is the linear growth rate, defined as
\begin{equation}
    f \equiv \frac{d\ln D_{1}}{d\ln a}.
\end{equation}
It is useful to rewrite Eqs.~(\ref{eq:delta_k})--(\ref{eq:theta_k}) as a single compact equation by introducing the doublet
\begin{equation}\label{eq:doublet}
    \Psi_a(\boldsymbol{k},T)
    =
    \left(
    \delta(\boldsymbol{k},T),
    -\frac{\theta(\boldsymbol{k},T)}{\mathcal{H}f}
    \right),
    \qquad
    T \equiv \ln D_{1},
\end{equation}
with the equations of motion now reading:
\begin{equation}
    \partial_{T}\Psi_a(\boldsymbol{k},T)
    + \Omega_{ab}\Psi_b(\boldsymbol{k},T)
    =
    \int_{\boldsymbol{k}_{1},\boldsymbol{k}_{2}}
    \delta^{(3)}_{\rm D}(\boldsymbol{k}_{1}+\boldsymbol{k}_{2}-\boldsymbol{k})
    \gamma_{abc}(\boldsymbol{k},\boldsymbol{k}_{1},\boldsymbol{k}_{2})
    \Psi_b(\boldsymbol{k}_{1},T)
    \Psi_c(\boldsymbol{k}_{2},T),
\end{equation}
with the mixing matrix:
\begin{equation}\label{eq:Omega_ab_EdS}
    \Omega_{ab}
    =
    \begin{pmatrix}
    0 & -1 \\
    -\frac{3}{2}\frac{\Omega_m}{f^2} &
    \frac{3}{2}\frac{\Omega_m}{f^2} - 1
    \end{pmatrix}.
\end{equation}
The non-zero components of the vertex function, $\gamma_{abc}$, are
\begin{equation}
    \gamma_{121} = \alpha(\boldsymbol{k}_{1},\boldsymbol{k}_{2}),
    \qquad
    \gamma_{222} = \beta(\boldsymbol{k}_{1},\boldsymbol{k}_{2}) .
\end{equation}

The assumption that $F_n$ and $G_n$ are separable in scale and time is exact in an Einstein--de Sitter universe, for which $\Omega_m=1$ and $f=1$. In this case, $\Omega_{ab}$ becomes time independent:
\begin{equation}
    \Omega_{ab}
    =
    \begin{pmatrix}
    0 & -1 \\
    -\frac{3}{2} & \frac{1}{2}
    \end{pmatrix}.
\end{equation}
Even though $\Omega_{\rm m}$ and $f$ differ from unity during dark-energy-dominated era, the ratio $\Omega_m/f^2$ remains close to unity in $\Lambda$CDM. This makes the Einstein--de Sitter approximation very accurate for $\Lambda$CDM.

Under the EdS approximation, Eqs.~(\ref{eq:delta_theta_n}) are written as:
\begin{align}\label{ref:delta_theta_n_EdS}
    \delta^{(n)}(\boldsymbol{k},\tau)
    &=D_{1}^{n}(\tau)
    \int_{\boldsymbol{k}_{1} \dots \boldsymbol{k}_{n}}
    \delta_{\rm D}^{(3)}
    (\boldsymbol{k}_{1}+ \dots + \boldsymbol{k}_{n} - \boldsymbol{k})
    F_{n}(\boldsymbol{k}_{1}, \dots, \boldsymbol{k}_{n})
    \left[
    \delta_{\rm lin}(\boldsymbol{k}_{1})
    \dots
    \delta_{\rm lin}(\boldsymbol{k}_{n})
    \right],
    \\
    \theta^{(n)}(\boldsymbol{k},\tau)
    &= - aHfD_{1}^{n}(\tau)
    \int_{\boldsymbol{k}_{1} \dots \boldsymbol{k}_{n}}
    \delta_{\rm D}^{(3)}
    (\boldsymbol{k}_{1}+ \dots + \boldsymbol{k}_{n} - \boldsymbol{k})
    G_{n}(\boldsymbol{k}_{1}, \dots, \boldsymbol{k}_{n})
    \left[
    \delta_{\rm lin}(\boldsymbol{k}_{1})
    \dots
    \delta_{\rm lin}(\boldsymbol{k}_{n})
    \right],
\end{align}
and the kernels up to second order are given by:
\begin{align}
    F_{1}(\boldsymbol{k})
    &=
    G_{1}(\boldsymbol{k})
    =
    1,
    \\
    F_{2}(\boldsymbol{k}_{1}, \boldsymbol{k}_{2})
    &=
    \frac{5}{7}\alpha_s(\boldsymbol{k}_{1}, \boldsymbol{k}_{2})
    +
    \frac{2}{7}\beta(\boldsymbol{k}_{1}, \boldsymbol{k}_{2}),
    \\
    G_{2}(\boldsymbol{k}_{1}, \boldsymbol{k}_{2})
    &=
    \frac{3}{7}\alpha_s(\boldsymbol{k}_{1}, \boldsymbol{k}_{2})
    +
    \frac{4}{7}\beta(\boldsymbol{k}_{1}, \boldsymbol{k}_{2}),
\end{align}
where $\alpha_s$ denotes the symmetrized version of the $\alpha$ kernel. The full recursive relations for the $n$-th order kernels $F_n$ and $G_n$ can be found in Ref.~\cite{Bernardeau:2001qr}.

This procedure provides a compact way to obtain the equal-time loop-corrected matter power spectrum, assuming Gaussian initial conditions, up to any desired perturbative order. At one-loop order we have:
\begin{equation}
    P^{\rm SPT}_{\rm 1-loop}(\boldsymbol{k}, \tau)
    =
    D^{2}_{1}(\tau)P_{\rm lin}(\boldsymbol{k})
    +
    D^{4}_{1}(\tau)
    \left[
    2P_{13}(\boldsymbol{k}) + P_{22}(\boldsymbol{k})
    \right],
\end{equation}
where $(2\pi)^{3}\delta^{\mathrm{D}}(\boldsymbol{k}-\boldsymbol{k}')P_{ij}(\boldsymbol{k})=\langle\delta^{(i)}(\boldsymbol{k})\delta^{(j)}(\boldsymbol{k}')\rangle$, and $i$ and $j$ are the respective order of the solution. This expression shows the decomposition of the matter power spectrum into the linear contribution and the leading nonlinear corrections, which arise from terms up to third order in the density field.

\section{Lagrangian Perturbation Theory}\label{sec:lpt}

So far we have discussed how to obtain 1-loop corrections to the matter power spectrum in the Eulerian description, where the density and velocity divergence fields describe the flow of the CDM fluid across fixed spatial coordinates. In this work, however, we follow Refs.~\cite{Matsu1,CLEFT,White:2014gfa,Vlah:2014nta} and employ a Lagrangian description of dark matter dynamics. In the Lagrangian formulation~\cite{Bouchet:1994xp,Zeldovich:1969sb,Matsubara:2015ipa,Buchert:1989xx,Moutarde:1991evx,Hivon:1994qb}, one follows the trajectories of fluid elements through the displacement field, $\boldsymbol{\Psi}(\boldsymbol{q},\tau)$. The starting point of this approach is the mapping between the Eulerian position of a particle, $\boldsymbol{x}(\tau)$, and its initial Lagrangian position, $\boldsymbol{q}$:
\begin{equation}\label{eq:lpt_master}
    \boldsymbol{x}(\tau)
    =
    \boldsymbol{q}
    +
    \boldsymbol{\Psi}(\boldsymbol{q},\tau),
\end{equation}
where the nonlinear displacement field $\boldsymbol{\Psi}(\boldsymbol{q},\tau)$ fully determines the evolution from the initial to the final particle position. For a non-relativistic self-gravitating fluid element, or particle, the geodesic equation governing its trajectory is:
\begin{equation}\label{eq:geo_eq}
    \frac{d^{2}\boldsymbol{x}}{d\tau^{2}}
    +
    \mathcal{H}\frac{d\boldsymbol{x}}{d\tau}
    =
    -
    \boldsymbol{\nabla}_{\boldsymbol{x}}\psi(\boldsymbol{x},\tau),
\end{equation}
where $\psi(\boldsymbol{x},\tau)$ is the Newtonian gravitational potential and $\boldsymbol{\nabla}_{\boldsymbol{x}}$ denotes the gradient operator with respect to Eulerian coordinates. Taking the divergence of Eq.~(\ref{eq:geo_eq}) and combining it with Eq.~(\ref{eq:lpt_master}), we obtain
\begin{equation}\label{eq:Psi_eq_LPT}
    \nabla_{\boldsymbol{x}}
    \cdot
    \left[
        \frac{d^{2}\boldsymbol{\Psi}}{d\tau^{2}}
        +
        \mathcal{H}\frac{d\boldsymbol{\Psi}}{d\tau}
    \right]
    =
    -
    4\pi G\bar{\rho}\,
    \delta(\boldsymbol{x},\tau).
\end{equation}

Mass conservation further implies that
\begin{equation}\label{eq:mass_cons}
    \bar{\rho}
    \left[
        1+\delta(\boldsymbol{x},\tau)
    \right]
    d^{3}\boldsymbol{x}
    =
    \bar{\rho}\,
    d^{3}\boldsymbol{q},
\end{equation}
which introduces the Jacobian of the Lagrangian-to-Eulerian mapping:
\begin{equation}
    J(\boldsymbol{q},\tau)
    =
    \left|
        \frac{\partial\boldsymbol{x}}
        {\partial\boldsymbol{q}}
    \right|.
\end{equation}
This leads to a relation between the density field and the Jacobian of the coordinate transformation:
\begin{equation}\label{eq:delta_jacob}
    1+\delta(\boldsymbol{x},\tau)
    =
    \frac{1}{J(\boldsymbol{q},\tau)}.
\end{equation}
An alternative form of Eq.~(\ref{eq:Psi_eq_LPT}) written in Lagrangian coordinates is:
\begin{equation}\label{eq:psi_jacob}
    (J^{-1})_{ij}
    \nabla_{q_j}
    \left[
        \frac{d^{2}\Psi_i}{d\tau^{2}}
        +
        \mathcal{H}\frac{d\Psi_i}{d\tau}
    \right]
    =
    -
    4\pi G\bar{\rho}\,
    \frac{
        1-J(\boldsymbol{q},\tau)
    }{
        J(\boldsymbol{q},\tau)
    },
\end{equation}
and the Jacobian matrix is:
\begin{equation}\label{eq:jacob_def}
    J_{ij}
    =
    \delta_{ij}
    +
    \Psi_{i,j}(\boldsymbol{q},\tau),
\end{equation}
where
\begin{equation}
    \Psi_{i,j}
    \equiv
    \frac{\partial\Psi_i}{\partial q_j}.
\end{equation}
Much like in the Eulerian prescription, we can also develop a perturbative treatment in the Lagrangian picture, now in terms of the displacement field, $\boldsymbol{\Psi}$. To do so, we begin by expanding the inverse Jacobian matrix:
\begin{equation}\label{eq:jacob_inv}
    (J^{-1})_{ij}
    =
    \delta_{ij}
    -
    \Psi_{i,j}
    +
    \Psi_{i,k}\Psi_{k,j}
    + \dots
\end{equation}
and by using the identities
\begin{align}\label{eq:jacob_exp}
    &J = 1 + J_{1} + J_{2} + J_{3}, \\
    &1 - \frac{1}{J}
    =
    J_{1}
    +
    J_{2}
    -
    J_{1}^{2}
    +
    J_{3}
    -
    2J_{1}J_{2}
    +
    J_{1}^{3}
    + \dots,
\end{align}
we arrive at the fully symmetrized contributions to the Jacobian expansion\footnote{To derive these equations, one only needs
\begin{align*}
    J
    =
    \frac{1}{6}
    \epsilon_{ijk}
    \epsilon_{pqr}
    J_{ip}J_{jq}J_{kr},
\end{align*}
together with Eq.~(\ref{eq:jacob_def}).}:
\begin{align}\label{eq:jacob_psis}
    J_{1}
    &=
    \Psi_{i,i},
    \nonumber\\
    J_{2}
    &=
    \frac{1}{2}
    \left[
        (\Psi_{i,i})^{2}
        -
        \Psi_{i,j}\Psi_{j,i}
    \right],
    \nonumber\\
    J_{3}
    &=
    \frac{1}{6}
    (\Psi_{i,i})^{3}
    -
    \frac{1}{2}
    \Psi_{i,i}\Psi_{j,k}\Psi_{k,j}
    +
    \frac{1}{3}
    \Psi_{i,j}\Psi_{j,k}\Psi_{k,i}.
\end{align}

Using Eq.~(\ref{eq:delta_jacob}), we may also write the matter overdensity in terms of the displacement field expansion:
\begin{align}\label{eq:delta_psis}
    \delta(\boldsymbol{x})
    =
    &
    -\Psi_{i,i}
    +
    \frac{1}{2}
    \left[
        (\Psi_{i,i})^{2}
        -
        \Psi_{i,j}\Psi_{j,i}
    \right]
    \nonumber\\
    &
    -
    \frac{1}{6}
    (\Psi_{i,i})^{3}
    -
    \frac{1}{3}
    \Psi_{i,j}\Psi_{j,k}\Psi_{k,i}
    +
    \frac{1}{2}
    \Psi_{i,i}\Psi_{j,k}\Psi_{k,j}
    +
    \mathcal{O}(\lambda^{4}).
\end{align}

Combining Eqs.~(\ref{eq:jacob_inv}), (\ref{eq:jacob_exp}), and (\ref{eq:delta_psis}), we may write the Fourier transform of Eq.~(\ref{eq:psi_jacob}) as
\begin{align}\label{eq:master_fourier_psi}
    \left[
        \hat{T}
        -
        \kappa
    \right]
    \Psi_{i,i}(\boldsymbol{k})
    =
    &
    \left[
        \Psi_{i,j}\hat{T}\Psi_{j,i}
    \right](\boldsymbol{k})
    -
    \frac{\kappa}{2}
    \left[
        \Psi_{i,j}\Psi_{j,i}
        -
        (\Psi_{i,i})^{2}
    \right](\boldsymbol{k})
    \nonumber\\
    &
    -
    \left[
        \Psi_{i,k}\Psi_{k,j}\hat{T}\Psi_{j,i}
    \right](\boldsymbol{k})
    +
    \frac{\kappa}{6}
    \left[
        (\Psi_{i,i})^{3}
    \right](\boldsymbol{k})
    \nonumber\\
    &
    +
    \frac{\kappa}{2}
    \left[
        \Psi_{i,j}\Psi_{j,i}\Psi_{k,k}
    \right](\boldsymbol{k})
    -
    \frac{\kappa}{3}
    \left[
        \Psi_{i,k}\Psi_{k,j}\Psi_{j,i}
    \right](\boldsymbol{k}).
\end{align}
where
\begin{equation}
    \kappa
    =
    4\pi G\bar{\rho},
\end{equation}
and, for brevity, we defined the differential operator
\begin{equation}
    \hat{T}
    =
    \frac{d^{2}}{d\tau^{2}}
    +
    \mathcal{H}\frac{d}{d\tau}.
\end{equation}
Writing Eq.~(\ref{eq:master_fourier_psi}) in this form eases the derivation of the equations of motion order by order in perturbation theory. The displacement field can then be expanded as
\begin{equation}\label{eq:Psi_expansion}
    \boldsymbol{\Psi}
    =
    \sum_{n=1}^{\infty}
    \boldsymbol{\Psi}^{(n)}
    =
    \boldsymbol{\Psi}^{(1)}
    +
    \boldsymbol{\Psi}^{(2)}
    +
    \boldsymbol{\Psi}^{(3)}
    + \dots.
\end{equation}

According to the Helmholtz decomposition theorem, a general vector field can be separated into longitudinal and transverse components. The displacement field may thus be written as
\begin{equation}
    \boldsymbol{\Psi}
    =
    \boldsymbol{\Psi}_{\rm L}
    +
    \boldsymbol{\Psi}_{\rm T},
\end{equation}
with
\begin{equation}
    \boldsymbol{\nabla}\times\boldsymbol{\Psi}_{\rm L}
    =
    0,
    \qquad
    \boldsymbol{\nabla}\cdot\boldsymbol{\Psi}_{\rm T}
    =
    0.
\end{equation}
Although vorticity is not generated perturbatively in the Eulerian description of a pressureless fluid prior to shell crossing, in LPT a transverse component naturally emerges in the displacement field~\cite{Matsubara:2015ipa,Rampf:2012up,Bouchet:1994xp,Buchert:1987xy}. In fact, solving the perturbative equations consistently up to third order already generates a non-vanishing transverse contribution. This component is required to ensure that the Eulerian velocity field remains irrotational before shell crossing.

%\subsection{1-loop matter power spectrum from LPT}

Before proceeding to the computation of the Eulerian-space power spectrum, it is important to briefly discuss the role of infrared (IR) resummation in large-scale structure perturbation theory. The need for infrared resummation arises because long-wavelength perturbations with momenta $p \lesssim k_{\rm BAO}$ induce coherent large-scale displacements that shift small-scale perturbations with $k \gg k_{\rm BAO}$. As a consequence, a naive perturbative expansion leads to a poor description of the BAO feature, since the linear BAO oscillations become progressively non-correlated from the higher-order corrected power spectrum. The IR-resummation procedure is therefore designed to resum only the enhanced long-wavelength displacement modes responsible for this effect. In Standard Perturbation Theory (SPT), this limitation of naive perturbative expansions has long been recognized and addressed in several works~\cite{Senatore:2014via,Senatore:2017pbn,LEFT,Baldauf:2015xfa,Vlah:2015zda,Blas:2016sfa,Ding:2017gad,Ivanov:2018gjr}. 

Obtaining the observable Eulerian-space power spectrum from LPT is, however, not straightforward. But, one of the main advantages of the approach developed in Refs.~\cite{Taylor:1996ne,Matsu1} is the natural resummation of modes enhanced by long-wavelength displacement effects. Within LPT, this issue is naturally alleviated because the formalism is constructed directly in terms of Lagrangian coordinates, $\boldsymbol{q}$. In particular, the Zel'dovich approximation (ZA) already captures the leading nonlinear coherent bulk flows of matter, effectively resumming the dominant large-scale displacement effects from the outset. Properly accounting for these large-scale motions is therefore essential for obtaining accurate predictions beyond linear order and for preserving the perturbative description of the BAO feature.

To compute the loop-corrected two-point statistics of matter in
Fourier space, we go back to the mass conservation relation, (\ref{eq:mass_cons}), but now written as:
\begin{equation}
    \rho(\boldsymbol{x},\tau)\,
    d^{3}\boldsymbol{x}
    =
    \bar{\rho}(\tau)\,
    d^{3}\boldsymbol{q},
\end{equation}
which can be recast as:
\begin{equation}\label{eq:mass_conserv_lpt}
    1
    +
    \delta(\boldsymbol{x},\tau)
    =
    \int
    d^{3}\boldsymbol{q}\;
    \delta^{\rm D}
    \!\left(
        \boldsymbol{x}
        -
        \boldsymbol{q}
        -
        \boldsymbol{\Psi}(\boldsymbol{q},\tau)
    \right),
\end{equation}
where $\delta^{\rm D}$ denotes the Dirac delta distribution enforcing the
mapping from Lagrangian to Eulerian coordinates. In Fourier space\footnote{Throughout this work, we adopt the Fourier convention
\begin{align*}
    F(\boldsymbol{x})
    &=
    \int
    \frac{d^{3}\boldsymbol{k}}{(2\pi)^{3}}
    F(\boldsymbol{k})
    e^{i\boldsymbol{k}\cdot\boldsymbol{x}},
    \\
    F(\boldsymbol{k})
    &=
    \int
    d^{3}\boldsymbol{x}\,
    F(\boldsymbol{x})
    e^{-i\boldsymbol{k}\cdot\boldsymbol{x}}.
\end{align*}} Eq.~(\ref{eq:mass_conserv_lpt}) becomes
\begin{equation}
    \delta(\boldsymbol{k},\tau)
    =
    \int
    d^{3}\boldsymbol{q}\,
    e^{-i\boldsymbol{k}\cdot\boldsymbol{q}}
    \left[
        e^{-i\boldsymbol{k}\cdot\boldsymbol{\Psi}(\boldsymbol{q},\tau)}
        -
        1
    \right],
\end{equation}
and, assuming statistical homogeneity, the matter power spectrum is:
\begin{equation}
    (2\pi)^{3}
    \delta^{\rm D}(\boldsymbol{k})
    +
    P_{\rm LPT}(k)
    =
    \int
    d^{3}\boldsymbol{q}\,
    e^{-i\boldsymbol{k}\cdot\boldsymbol{q}}
    \left\langle
        e^{-i\boldsymbol{k}\cdot
        \Delta\boldsymbol{\Psi}(\boldsymbol{q},\tau)}
    \right\rangle,
\end{equation}
where
\begin{equation}
    \Delta\boldsymbol{\Psi}(\boldsymbol{q},\tau)
    \equiv
    \boldsymbol{\Psi}(\boldsymbol{q}_{1},\tau)
    -
    \boldsymbol{\Psi}(\boldsymbol{q}_{2},\tau),
\end{equation}
with
\begin{equation}
    \boldsymbol{q}
    \equiv
    \boldsymbol{q}_{1}
    -
    \boldsymbol{q}_{2}.
\end{equation}

Using the cumulant theorem~\cite{Matsu1},
\begin{equation}
    \left\langle
        e^{-iX}
    \right\rangle
    =
    \exp
    \left[
        \sum_{N=1}^{\infty}
        \frac{(-i)^{N}}{N!}
        \left\langle
            X^{N}
        \right\rangle_{c}
    \right],
\end{equation}
where $\langle X^{N}\rangle_{c}$ denotes the $N$-th cumulant of the random variable $X$~\cite{Bernardeau:2001qr}, one rewrites the power spectrum as
\begin{equation}\label{eq:pk_lpt_cdm}
    (2\pi)^{3}
    \delta^{\rm D}(\boldsymbol{k})
    +
    P_{\rm LPT}(k)
    =
    \int
    d^{3}\boldsymbol{q}\,
    e^{-i\boldsymbol{k}\cdot\boldsymbol{q}}
    \exp
    \left[
        -\frac{1}{2}
        k_{i}k_{j}
        A_{ij}(\boldsymbol{q})
        +
        \frac{i}{6}
        k_{i}k_{j}k_{k}
        W_{ijk}(\boldsymbol{q})
    \right],
\end{equation}
where
\begin{equation}
    A_{ij}(\boldsymbol{q})
    \equiv
    \left\langle
        \Delta\Psi_{i}
        \Delta\Psi_{j}
    \right\rangle_{c},
\end{equation}
and
\begin{equation}
    W_{ijk}(\boldsymbol{q})
    \equiv
    \left\langle
        \Delta\Psi_{i}
        \Delta\Psi_{j}
        \Delta\Psi_{k}
    \right\rangle_{c}.
\end{equation}
Explicitly~\cite{Matsu1,CLEFT,LEFT},
\begin{align}\label{eq:A_ij_W_ijk}
    A_{ij}(\boldsymbol{q})
    &=
    2
    \left[
        \left\langle
            \Psi_{i}(\boldsymbol{0})
            \Psi_{j}(\boldsymbol{0})
        \right\rangle
        -
        \left\langle
            \Psi_{i}(\boldsymbol{q_{1}})
            \Psi_{j}(\boldsymbol{q_{2}})
        \right\rangle
    \right],
    \nonumber\\
    W_{ijk}(\boldsymbol{q})
    &=
    \left\langle
        \Psi_{\{i}(\boldsymbol{q}_{1})
        \Psi_{j}(\boldsymbol{q}_{1})
        \Psi_{k\}}(\boldsymbol{q}_{2})
    \right\rangle
    -
    \left\langle
        \Psi_{\{i}(\boldsymbol{q}_{2})
        \Psi_{j}(\boldsymbol{q}_{2})
        \Psi_{k\}}(\boldsymbol{q}_{1})
    \right\rangle,
\end{align}
where $\{\cdots\}$ denotes symmetrization over the enclosed indices and $\boldsymbol{q}=\boldsymbol{q}_{1} - \boldsymbol{q}_{}$.

Perturbation theory in the displacement field, Eq.~(\ref{eq:Psi_expansion}), then yields expressions for $\boldsymbol{A}$ and $\boldsymbol{W}$ in terms of momentum-dependent integrals. For example,
\begin{equation}\label{eq:A_ij_lin_loop}
    A_{ij}
    =
    A_{ij}^{\rm lin}
    +
    A_{ij}^{(22)}
    +
    2A_{ij}^{(13)},
\end{equation}
where the linear contribution corresponds to the Zel'dovich approximation, i.e.\ $A_{ij}^{\rm lin}\equiv A_{ij}^{(11)}$, and the symmetry relation $(13)=(31)$ has been used.

More generally, it is useful to define the Lagrangian correlator polyspectra~\cite{Matsu1,CLEFT,Aviles:2017aor,LEFT},
\begin{equation}\label{eq:lag_corr_poly}
    C_{i_{1}i_{2}\dots}^{(m_{1}m_{2}\dots)}
    \equiv
    \left\langle
        \Delta\Psi_{i_{1}}^{(m_{1})}
        \Delta\Psi_{i_{2}}^{(m_{2})}
        \dots
    \right\rangle,
\end{equation}
which, for the pure matter case, depend solely on the displacement fields. Later, we will generalize these correlators to the case of biased tracers. At 1-loop order, the relevant correlators are
\begin{align}\label{eq:lag_corr_examples}
    C_{ij}^{(11)}(\boldsymbol{k})
    &=
    L_{i}^{(1)}(\boldsymbol{k})
    L_{j}^{(1)}(\boldsymbol{k})
    P_{\rm L}(k),
    \nonumber\\
    C_{ij}^{(22)}(\boldsymbol{k})
    &=
    \frac{1}{2}
    \int
    \frac{d^{3}\boldsymbol{p}}{(2\pi)^{3}}
    L_{i}^{(2)}
    (\boldsymbol{p},\boldsymbol{k-p})
    L_{j}^{(2)}
    (\boldsymbol{p},\boldsymbol{k-p})
    P_{\rm L}(p)
    P_{\rm L}(|\boldsymbol{k-p}|),
    \nonumber\\
    C_{ij}^{(13)}(\boldsymbol{k})
    &=
    \frac{1}{2}
    L_{i}^{(1)}(\boldsymbol{k})
    P_{\rm L}(k)
    \int
    \frac{d^{3}\boldsymbol{p}}{(2\pi)^{3}}
    L_{j}^{(3,\mathrm{symm.})}
    (\boldsymbol{k},-\boldsymbol{p},\boldsymbol{p})
    P_{\rm L}(p),
    \nonumber\\
    C_{ijk}^{(112)}
    (\boldsymbol{k}_{1},\boldsymbol{k}_{2},\boldsymbol{k}_{3})
    &=
    -
    L_{i}^{(1)}(\boldsymbol{k}_{1})
    L_{j}^{(1)}(\boldsymbol{k}_{2})
    L_{k}^{(2)}
    (\boldsymbol{k}_{1},\boldsymbol{k}_{2})
    P_{\rm L}(k_{1})
    P_{\rm L}(k_{2}).
\end{align}
Here,
$
\boldsymbol{L}^{(1)},
\boldsymbol{L}^{(2)},
$
and
$
\boldsymbol{L}^{(3,\mathrm{symm.})}
$
denote the first-, second-, and symmetrized third-order Lagrangian kernels, respectively. Recall that the displacement field develops a transverse component at third order. However, at 1-loop order only the symmetric longitudinal contribution enters, which considerably simplifies the calculations. Formally, for general cosmologies, the Lagrangian kernels may be written in terms of generalized growth functions, which in general acquire nontrivial momentum dependence beyond the Einstein--de Sitter (EdS) approximation:
\begin{align}\label{eq:lag_kernels_gen}
    \boldsymbol{L}^{(1)}(\boldsymbol{k})
    &=
    \frac{\boldsymbol{k}}{k^{2}},
    \\
    \boldsymbol{L}^{(2)}
    (\boldsymbol{k}_{1},\boldsymbol{k}_{2})
    &=
    \frac{\boldsymbol{k}}{k^{2}}
    \frac{
        D^{(2)}
        (\boldsymbol{k}_{1},\boldsymbol{k}_{2})
    }{
        D^{(1)}(k_{1})
        D^{(1)}(k_{2})
    },
    \\
    \boldsymbol{L}^{(3,\mathrm{symm.})}
    (\boldsymbol{k}_{1},\boldsymbol{k}_{2},\boldsymbol{k}_{3})
    &=
    \frac{\boldsymbol{k}}{k^{2}}
    \frac{
        D^{(3),\mathrm{symm.}}
        (\boldsymbol{k}_{1},\boldsymbol{k}_{2},\boldsymbol{k}_{3})
    }{
        D^{(1)}(k_{1})
        D^{(1)}(k_{2})
        D^{(1)}(k_{3})
    }.
\end{align}
In beyond EdS, the generalized growth functions depend not only on time but also on the momentum configuration, thereby encoding the angular dependence of the kernels.

For Einstein--de Sitter cosmologies, these kernels assume fixed analytical forms:
\begin{align}
    \boldsymbol{L}_{\rm EdS}^{(1)}(\boldsymbol{k})
    &=
    \frac{\boldsymbol{k}}{k^2},
    \\
    \boldsymbol{L}_{\rm EdS}^{(2)}
    (\boldsymbol{k}_1,\boldsymbol{k}_2)
    &=
    \frac{3}{7}\,
    \frac{\boldsymbol{k}}{k^2}
    \left[
        1
        -
        \left(
            \frac{
                \boldsymbol{k}_1\cdot\boldsymbol{k}_2
            }{
                k_1 k_2
            }
        \right)^2
    \right],
    \\
    \boldsymbol{L}_{\rm EdS}^{(3,\mathrm{symm.})}
    (\boldsymbol{k}_1,\boldsymbol{k}_2,\boldsymbol{k}_3)
    &=
    \frac{5}{7}\,
    \frac{\boldsymbol{k}}{k^2}
    \left[
        1
        -
        \left(
            \frac{
                \boldsymbol{k}_1\cdot\boldsymbol{k}_2
            }{
                k_1 k_2
            }
        \right)^2
    \right]
    \left\{
        1
        -
        \left[
            \frac{
                (\boldsymbol{k}_1+\boldsymbol{k}_2)
                \cdot
                \boldsymbol{k}_3
            }{
                |\boldsymbol{k}_1+\boldsymbol{k}_2|\,k_3
            }
        \right]^2
    \right\}
    \nonumber\\
    &\quad
    -
    \frac{1}{3}\,
    \frac{\boldsymbol{k}}{k^2}
    \left[
        1
        -
        3
        \left(
            \frac{
                \boldsymbol{k}_1\cdot\boldsymbol{k}_2
            }{
                k_1 k_2
            }
        \right)^2
        +
        2
        \frac{
            (\boldsymbol{k}_1\cdot\boldsymbol{k}_2)
            (\boldsymbol{k}_2\cdot\boldsymbol{k}_3)
            (\boldsymbol{k}_3\cdot\boldsymbol{k}_1)
        }{
            k_1^2 k_2^2 k_3^2
        }
    \right].
\end{align}
These EdS-kernels are fully time independent and possess fixed momentum dependence, which allows them to be directly substituted into the perturbative expansion of the displacement field:
\begin{align}\label{eq:Psi_n_EdS}
    \boldsymbol{\Psi}_{\rm EdS}^{(n)}
    (\boldsymbol{k},a)
    =
    i\,D^{n}(a)
    \int
    \frac{d^3\boldsymbol{k}_1}{(2\pi)^3}
    \cdots
    \frac{d^3\boldsymbol{k}_n}{(2\pi)^3}
    \,
    (2\pi)^3
    \delta^{\rm D}
    \!\left(
        \boldsymbol{k}
        -
        \sum_{i=1}^{n}
        \boldsymbol{k}_i
    \right)
    \nonumber\\
    \times
    \boldsymbol{L}_{\rm EdS}^{(n)}
    (\boldsymbol{k}_1,\dots,\boldsymbol{k}_n)
    \,
    \delta_0(\boldsymbol{k}_1)
    \cdots
    \delta_0(\boldsymbol{k}_n).
\end{align}
Importantly, the Lagrangian correlators are not restricted to cosmologies well approximated by EdS. One may instead compute the generalized growth functions up to the symmetric third-order contribution and construct the corresponding 1-loop matter power spectrum for more general cosmological models, and Equation~(\ref{eq:Psi_n_EdS}) becomes:
\begin{align}
    \boldsymbol{\Psi}^{(n)}
    (\boldsymbol{k},a)
    =
    i
    \int
    \frac{d^3\boldsymbol{k}_1}{(2\pi)^3}
    \cdots
    \frac{d^3\boldsymbol{k}_n}{(2\pi)^3}
    \,
    (2\pi)^3
    \delta^{\rm D}
    \!\left(
        \boldsymbol{k}
        -
        \sum_{i=1}^{n}
        \boldsymbol{k}_i
    \right)
    \nonumber\\
    \times
    \boldsymbol{L}^{(n)}
    (\boldsymbol{k}_1,\dots,\boldsymbol{k}_n;a)
    \,
    \delta_0(\boldsymbol{k}_1)
    \cdots
    \delta_0(\boldsymbol{k}_n).
\end{align}
From the definition of the Lagrangian correlators, it is useful to define the following scalar functions, following Refs.~\cite{Matsu1,LEFT,Aviles:2017aor,Vlah:2016bcl}:
\begin{align}
    Q_1(k)
    &=
    \frac{98}{9}\,
    k_i k_j\,
    C^{(22)}_{ij}(k),
    \label{eq:Q1_lag}
    \\
    Q_2(k)
    &=
    \frac{7}{3}\,
    k_i k_j k_k
    \int
    \frac{d^3\boldsymbol{p}}{(2\pi)^3}\,
    C^{(211)}_{ijk}
    (\boldsymbol{k},-\boldsymbol{p},\boldsymbol{p}-\boldsymbol{k}),
    \label{eq:Q2_lag}
    \\
    Q_3(k)
    &=
    k_i k_j k_k k_l
    \int
    \frac{d^3\boldsymbol{p}}{(2\pi)^3}
    \,
    C^{(11)}_{ij}(\boldsymbol{p})
    C^{(11)}_{kl}(\boldsymbol{k}-\boldsymbol{p}),
    \label{eq:Q3_lag}
    \\
    R_1(k)
    &=
    \frac{21}{5}\,
    k_i k_j\,
    C^{(13)}_{ij}(k),
    \label{eq:R1_lag}
    \\
    R_2(k)
    &=
    \frac{7}{3}\,
    k_i k_j k_k
    \int
    \frac{d^3\boldsymbol{p}}{(2\pi)^3}\,
    C^{(112)}_{ijk}
    (\boldsymbol{k},-\boldsymbol{p},\boldsymbol{p}-\boldsymbol{k})
    \nonumber\\
    &=
    \frac{7}{3}\,
    k_i k_j k_k
    \int
    \frac{d^3\boldsymbol{p}}{(2\pi)^3}\,
    C^{(121)}_{ijk}
    (\boldsymbol{k},-\boldsymbol{p},\boldsymbol{p}-\boldsymbol{k}).
    \label{eq:R2_lag}
\end{align}

Within the Einstein--de Sitter approximation, Eqs.~(\ref{eq:Q1_lag})--(\ref{eq:R2_lag}) simplify considerably. In particular, the $R_n$ functions may be written as
\begin{equation}
    R_n^{\rm EdS}(k)
    =
    \frac{k^3}{(2\pi)^2}
    P_{\rm L}(k)
    \int_0^\infty
    dr\,
    P_{\rm L}(kr)
    \,
    \widetilde{R}_n(r),
\end{equation}
where
\begin{align}
    \widetilde{R}_1(r)
    &=
    \int_{-1}^{1}
    d\mu\,
    \frac{
        r^2
        (1-\mu^2)^2
    }{
        1+r^2-2r\mu
    },
    \\
    \widetilde{R}_2(r)
    &=
    \int_{-1}^{1}
    d\mu\,
    \frac{
        (1-\mu^2)\,
        r\mu\,
        (1-r\mu)
    }{
        1+r^2-2r\mu
    }.
\end{align}
Similarly, the $Q_n$ functions become
\begin{equation}
    Q_n^{\rm EdS}(k)
    \equiv
    \frac{k^3}{(2\pi)^2}
    \int_0^\infty
    dr\,
    P_{\rm L}(kr)
    \int_{-1}^{+1}
    d\mu\,
    P_{\rm L}
    \!\left(
        k
        \sqrt{
            1+r^2-2r\mu
        }
    \right)
    \widetilde{Q}_n(r,\mu),
\end{equation}
with
\begin{align}
    \widetilde{Q}_1(r,\mu)
    &=
    \frac{
        r^2
        (1-\mu^2)^2
    }{
        (1+r^2-2r\mu)^2
    },
    \\
    \widetilde{Q}_2(r,\mu)
    &=
    \frac{
        (1-\mu^2)\,
        r\mu\,
        (1-r\mu)
    }{
        (1+r^2-2r\mu)^2
    },
    \\
    \widetilde{Q}_3(r,\mu)
    &=
    \frac{
        \mu^2
        (1-r\mu)^2
    }{
        (1+r^2-2r\mu)^2
    },
\end{align}
where these expressions have already been rewritten in spherical coordinates
\begin{equation}
    r
    \equiv
    \frac{p}{k},
    \qquad
    \mu
    \equiv
    \hat{\boldsymbol{k}}
    \cdot
    \hat{\boldsymbol{p}},
\end{equation}
and the generalized form of $R_n(k)$ and $Q_n(k)$, is found in~\cite{Aviles:2017aor}. 

A straightforward numerical approach consists of performing a Gauss--Legendre quadrature for the angular integration together with a trapezoidal integration for the radial part. Within the EdS approximation, however, these integrals can be computed much more efficiently, as shown in Ref.~\cite{Schmittfull:2016jsw}. The key observation is that these integrals possess a separable angular structure that can be decomposed into Legendre polynomials. In this case, the loop integrals may be recast as
\begin{equation}\label{eq:fast_integ}
    \int
    d^{3}\boldsymbol{p}\,
    p^{n_{1}}
    |\boldsymbol{k-p}|^{n_{2}}
    P_{\ell}
    \!\left(
        \hat{\boldsymbol{p}}
        \cdot
        \widehat{(\boldsymbol{k-p})}
    \right)
    P_{\rm L}(p)
    P_{\rm L}(|\boldsymbol{k-p}|)
    =
    (-1)^{\ell}
    4\pi
    \int
    dr\,
    r^{2}
    j_{0}(kr)
    \xi_{n_{1}}^{\ell}(r)
    \xi_{n_{2}}^{\ell}(r),
\end{equation}
where $P_{\ell}$ are the Legendre polynomials and $\xi_{n}^{\ell}(r)$ denotes the generalized linear correlation function:
\begin{align}
    \xi_{n}^{\ell}(r)
    &=
    i^{\ell}
    \int
    \frac{d^{3}\boldsymbol{k}}{(2\pi)^{3}}
    e^{-i\boldsymbol{k}\cdot\boldsymbol{r}}
    k^{n}
    P_{\ell}
    (\hat{\boldsymbol{k}}\cdot\hat{\boldsymbol{r}})
    P_{\rm L}(k)
    \nonumber\\
    &=
    i^{\ell}
    \int
    \frac{dk}{2\pi^{2}}
    k^{n+2}
    j_{\ell}(kr)
    P_{\rm L}(k).
\end{align}
The explicit expressions for the $Q_n$ and $R_n$ functions in terms of $\xi_n^\ell$ may be found in Appendix~B of Ref.~\cite{Schmittfull:2016jsw}. An important remark is that, for cosmologies in which the linear growth function becomes scale dependent, a finite-dimensional decomposition in terms of Legendre polynomials is generally no longer possible. This occurs because the functions $\widetilde{Q}_n$ and $\widetilde{R}_n$ no longer possess fixed analytical dependence on the angular and radial variables. Nevertheless, one may still follow the same general strategy introduced in Ref.~\cite{Schmittfull:2016jsw}. We present this generalized approach in Appendix~\ref{app:fast_integrals_mg}.

Returning to Eqs.~(\ref{eq:pk_lpt_cdm}) and (\ref{eq:A_ij_W_ijk}), we now follow the notation of Refs.~\cite{LEFT,Aviles:2017aor,Chen:2020fxs} and decompose the connected correlators of the Lagrangian displacement field into scalar functions constructed from $\hat{\boldsymbol{q}}$:
\begin{align}
    A_{ij}(\boldsymbol{q})
    &=
    X(q)\delta_{ij}
    +
    Y(q)\hat{q}_{i}\hat{q}_{j},
    \\
    W_{ijk}(\boldsymbol{q})
    &=
    V(q)\hat{q}_{\{i}\delta_{jk\}}
    +
    T(q)\hat{q}_{i}\hat{q}_{j}\hat{q}_{k}.
\end{align}
The scalar functions are given by
\begin{align}
    X(q)
    &=
    \frac{1}{\pi^{2}}
    \int_{0}^{\infty}
    dk
    \left[
        P_{\rm L}(k)
        +
        \frac{9}{98}Q_{1}(k)
        +
        \frac{10}{21}R_{1}(k)
    \right]
    \left(
        \frac{1}{3}
        -
        \frac{j_{1}(kq)}{kq}
    \right),\label{eq:X}
    \\
    Y(q)
    &=
    \frac{1}{\pi^{2}}
    \int_{0}^{\infty}
    dk
    \left[
        P_{\rm L}(k)
        +
        \frac{9}{98}Q_{1}(k)
        +
        \frac{10}{21}R_{1}(k)
    \right]
    j_{2}(kq),\label{eq:Y}
    \\
    T(q)
    &=
    -
    \frac{9}{14\pi^{2}}
    \int_{0}^{\infty}
    \frac{dk}{k}
    \left[
        Q_{1}(k)
        +
        2Q_{2}(k)
        +
        2R_{1}(k)
        +
        4R_{2}(k)
    \right]
    j_{3}(kq),\label{eq:T}
    \\
    V(q)
    &=
    -
    \frac{1}{7\pi^{2}}
    \int_{0}^{\infty}
    \frac{dk}{k}
    \left[
        Q_{1}(k)
        -
        3Q_{2}(k)
        +
        2R_{1}(k)
        -
        6R_{2}(k)
    \right]
    j_{1}(kq)
    -
    \frac{1}{5}T(q).\label{eq:V}
\end{align}

As can be seen in Eqs.~(\ref{eq:X}-\ref{eq:Y}), the functions $X(q)$ and $Y(q)$ naturally separate into linear and loop contributions. The linear pieces correspond to the terms containing only the linear power spectrum $P_{\rm L}(k)$:
\begin{align}
    X^{\rm lin}(q)
    &=
    \frac{1}{\pi^{2}}
    \int_{0}^{\infty}
    dk\,
    P_{\rm L}(k)
    \left(
        \frac{1}{3}
        -
        \frac{j_{1}(kq)}{kq}
    \right),
    \\
    Y^{\rm lin}(q)
    &=
    \frac{1}{\pi^{2}}
    \int_{0}^{\infty}
    dk\,
    P_{\rm L}(k)
    j_{2}(kq),
\end{align}
with the remaining contributions defining $X^{\rm loop}(q)$ and $Y^{\rm loop}(q)$.

Finally, to construct the 1-loop matter power spectrum, we adopt the convolution Lagrangian perturbation theory (CLPT) prescription introduced in Ref.~\cite{CLEFT,LEFT}. In this approach, the linear contribution in Eq.~(\ref{eq:pk_lpt_cdm}) is kept exponentiated, while the loop corrections are expanded perturbatively:
\begin{equation}
    P_{\rm LPT}^{\rm 1-loop}(k)
    =
    \int
    d^{3}\boldsymbol{q}\,
    e^{-i\boldsymbol{k}\cdot\boldsymbol{q}
    -
    \frac{1}{2}
    k_{i}k_{j}
    A_{ij}^{\rm lin}(\boldsymbol{q})}
    \left[
        1
        -
        \frac{1}{2}
        k_{i}k_{j}
        A_{ij}^{\rm loop}(\boldsymbol{q})
        +
        \frac{i}{6}
        k_{i}k_{j}k_{k}
        W_{ijk}(\boldsymbol{q})
        +
        \dots
    \right].
\end{equation}

Reference~\cite{LEFT} emphasizes the advantages of keeping the linear contribution exponentiated, as originally proposed in Ref.~\cite{Matsu1}. Within LPT, this procedure properly accounts for the advection induced by coherent large-scale bulk flows, effectively resumming infrared displacements non-perturbatively and improving the description of baryon acoustic oscillation features. Furthermore, in the spirit of an effective field theory treatment, perturbatively expanding the loop corrections avoids the uncontrolled proliferation of higher-order resummed counterterms, while maintaining a well-behaved connection between the power spectrum in Fourier space and the two-point correlation function in configuration space. This cleaner connection between Fourier and configuration space arises because the LPT power spectrum is already partially resummed through the exponentiation of displacement correlators, as reflected by the presence of the zero-lag contribution in the $\boldsymbol{A}$ term of Eq.~(\ref{eq:A_ij_W_ijk}). In particular, the correlation function depends directly on the relative large-scale displacements between particle pairs, whose variance naturally produces the smearing of the BAO feature. As a consequence, infrared bulk flows are incorporated non-perturbatively from the outset, leading to a more stable and physically transparent relation between configuration- and Fourier-space observables.

The formalism developed throughout this section can be straightforwardly generalized beyond cosmologies well described by the Einstein--de Sitter approximation. While we have presented several expressions explicitly in their EdS form, following the standard treatment in the literature, we have also introduced the corresponding generalized formulations valid for cosmologies with scale-dependent growth. These generalized expressions are the ones employed throughout this work when presenting our numerical results and assessing deviations from the EdS approximation.

In particular, we have developed a numerical implementation capable of solving the generalized growth functions up to third order, including the symmetrized contribution $D^{(3),\mathrm{symm.}}(\boldsymbol{k}_1,\boldsymbol{k}_2,\boldsymbol{k}_3;a)$, required to construct the Lagrangian kernels in scale-dependent modified gravity theories, as shown in Eq.~(\ref{eq:lag_kernels_gen}).

\section{Modified Gravity}\label{sec:mg}

The equations of motion that we solved have been carefully derived in~\cite{Aviles:2017aor,Aviles:2018qot} for scalar tensor theories where the scale-dependent growth naturally is endowed with the chameleon screening mechanism. The Lagrangian used has the following form:
\begin{equation}\label{eq:bd_lag}
    \mathcal{L} = \frac{1}{16\pi G}\sqrt{-g}\left[ \varphi R - \frac{\omega_{\rm BD}(\varphi)}{\varphi} \nabla_{\mu}\varphi \nabla^{\mu}\varphi - V(\varphi)\right] + \mathcal{L}_{\rm m}\left[\zeta, g_{\mu \nu}\right],
\end{equation}
where $\zeta$ are the matter fields minimally coupled to the metric. The Lagrangian (\ref{eq:bd_lag}) is known as a generalized formulation of the Brans-Dicke theory, where the coupling function $\omega_{\rm BD}(\phi)$ is a function of the scalar field, instead of a constant as in the original BD theory. This theory has long been studied, with many exact cosmological solutions being derived, as well as constrained using cosmological and astrophysical data.

Our interest here is not to pursue this theory as a viable candidate to a beyond-$\Lambda$CDM model, but to rather work out the details when computing the computing 1-loop corrections in a theory that its growth of structure is scale-dependent. To better understand this, we will show some of the equations of motion derived in~\cite{Aviles:2017aor} to get the third order growth factor in such theory, as well as shown intermediate definitions and derivations of a subclass of these theories, called Hu-Sawicki $f(R)$ model, that have also been studied in detail within SPT and 2LPT in~\cite{Koyama:2009me,Bose:2016qun,Winther:2017jof,Brando:2023fzu,Aviles:2017aor}. 

Using the perturbed FLRW metric:
\begin{equation}
    ds^2 = - (1+ 2\psi)dt^{2} + a^{2}(t)(1-2\phi)\delta_{ij}dx^{i}dx^{j},
\end{equation}
and considering fluctuations of the scalar field: $\varphi(\boldsymbol{x},t)\to\bar{\varphi}(t) + \delta \varphi(\boldsymbol{x},t)$ we arrive at~\cite{Koyama:2009me}:
\begin{align}
    &\frac{1}{a^{2}}\nabla^{2}\psi = 4\pi G \bar{\rho}\delta  - \frac{1}{2a^{2}} \nabla^{2}\delta\varphi,\label{eq:mg_poi}\\
    &\left(3+2\omega_{\rm BD}(\varphi)\right)\nabla^{2}\delta\varphi = - 8 \pi G \bar{\rho}\delta - I(\delta \varphi)\label{eq:mg_kg},
\end{align}
where
\begin{align}\label{eq:I_int}
I(\delta\varphi) &=\frac{1}{2} \int \frac{d^3 k_1\, d^3 k_2}{(2\pi)^3}\,
\delta^{\rm D}(\boldsymbol{k} - \boldsymbol{k}_{12})\,
M_2(\boldsymbol{k}_1,\boldsymbol{k}_2)\,\varphi(\boldsymbol{k}_1)\varphi(\boldsymbol{k}_2) \nonumber\\
&\quad + \frac{1}{6} \int \frac{d^3 k_1\, d^3 k_2\, d^3 k_3}{(2\pi)^6}\,
\delta^{\rm D}(\boldsymbol{k} - \boldsymbol{k}_{123})\,
M_3(\boldsymbol{k}_1,\boldsymbol{k}_2,\boldsymbol{k}_3)\,\varphi(\boldsymbol{k}_1)\varphi(\boldsymbol{k}_2)\varphi(\boldsymbol{k}_3)
+ \cdots
\end{align}
and to make the notation more compact we treat $\varphi$ now as $\delta \varphi$. The mass functions $M_{i}$ are in full generality time and scale-dependent, and are intended to incorporate screening mechanism corrections that shield the fictitious fifth force introduced by the scalar field in small scales. One can see from (\ref{eq:mg_poi}) and (\ref{eq:mg_kg}) that nonlinearities coming from the matter density field will source the fluctuations of the scalar field, and the $I(\varphi)$ function is the term that will be screening the presence of the scalar field correction to the gravitational force from appearing.

However, in perturbation theory, as we solve the continuity, Euler and now the Klein-Gordon equation iteratively starting from the linear solution, $\delta^{(1)}(\boldsymbol{x},t)$, it has been shown that this screening terms in not as ``effective", in the sense that its corrections are subdominant when compared to first term~\cite{Bose:2016qun,Rodriguez-Meza:2023rga}. Alternatively, one can check this by noting that the linearized Klein-Gordon equation:
\begin{equation}
    \nabla^{2}\varphi^{(1)} = - \frac{1}{2\omega_{\rm BD}+3}\kappa~\bar{\rho}\delta^{(1)}
\end{equation}
where we introduced $\kappa = 8 \pi G$, is already order unity in $\kappa$. Therefore, each time a $\varphi$ appears inside the integral in Eq.~(\ref{eq:I_int}), we get an additional order in $\kappa$, making it smaller. While in modified gravity simulations nonlinear terms in the Klein-Gordon equation, such as $I(\varphi)$, would also appear, and the scalar field will already contain a $\kappa$ factor in each term, the matter density field has no restriction in being linear, therefore, such nonlinear terms can always be of the same order as the first linear term in Eq.~(\ref{eq:mg_kg}).

We will now consider the Hu-Sawicki model, a special case of $f(R)$ theories of gravity. As it was shown~\cite{Sotiriou:2008rp,DeFelice:2010aj}, $f(R)$ theories in the metric formalism, i.e., when one varies the action with respect to the metric, $g_{\mu \nu}$, and not the connection term, can be mapped directly to the Brans-Dicke theory with $\omega_{\rm BD}=0$, therefore, the BD parameter drops out of the equations of motion. At the same time, at the background level we have that if:
\begin{equation}
    |f(R)| \ll 1, \ \ \left|\frac{f(R)}{\bar{R}}\right|\ll1
\end{equation}
with $f_{R} = df/dR$ one has an expansion history close to $\Lambda$CDM, thence, in this work the modified gravity results will always have a $\Lambda$CDM cosmology in the background. Furthermore, we have that the mass functions $M_{1}$, $M_{2}$ and $M_{3}$ are only time dependent:
\begin{align}
M_1(a) &= \frac{3}{2}\,\frac{H_0^2}{|f_{R0}|}
\frac{\left(\Omega_{m0}a^{-3} + 4\Omega_\Lambda\right)^3}
{\left(\Omega_{m0} + 4\Omega_\Lambda\right)^2}, \\
M_2(a) &= \frac{9}{4}\,\frac{H_0^2}{|f_{R0}|^2}
\frac{\left(\Omega_{m0}a^{-3} + 4\Omega_\Lambda\right)^5}
{\left(\Omega_{m0} + 4\Omega_\Lambda\right)^4}, \\
M_3(a) &= \frac{45}{8}\,\frac{H_0^2}{|f_{R0}|^3}
\frac{\left(\Omega_{m0}a^{-3} + 4\Omega_\Lambda\right)^7}
{\left(\Omega_{m0} + 4\Omega_\Lambda\right)^6}.
\end{align}

The full treatment and development of SPT in $f(R)$ gravity is presented in detail in~\cite{Koyama:2009me,Bose:2016qun}. Rather than reproducing the derivation here, we will present the form of the mixing matrix $\Omega_{ab}$, the same as Equation~(\ref{eq:Omega_ab_EdS}
), obtained when using the doublet representation of Equation~(\ref{eq:doublet}) in order to make a pedagogical remark regarding the EdS approximation and generalized cosmologies:
\begin{equation}\label{eq:Omega_ab_fr}
\Omega_{ab}(k,T)
=
\begin{pmatrix}
0 & -1 \\[10pt]
-\dfrac{3}{2}\dfrac{\Omega_m(a)}{f^2(a)}\,\mu(k,a)
&
1+\partial_T \ln\!\bigl[\mathcal{H}(a)\,f(a)\bigr]
\end{pmatrix},
\end{equation}
where
\begin{equation}
\mu(k,a)
=
1+\frac13\frac{(k/a)^2}{\Pi(k,a)}.
\end{equation}
Unlike the $\Lambda$CDM case, in generalized cosmologies in the EdS limit, where $\Omega_m/f^2 \rightarrow 1$, the modified gravity contribution $\mu(k,a)$ remains present in the component $\Omega_{21}$. This function enters directly in the Poisson equation, encoding the scale- and time-dependent modifications to gravity.

Returning to the Lagrangian Perturbation Theory approach, as in the $\Lambda$CDM case, the perturbative expansion of the displacement field yields the solution order by order. We begin by rewriting Equation~(\ref{eq:Psi_eq}), keeping the right-hand side explicitly in terms of the metric potential:
\begin{align}\label{eq:Psi_eq}
    \nabla_{\boldsymbol{x}}\cdot\left[
    \frac{d^{2}\boldsymbol{\Psi}}{d \tau^{2}}
    +
    \mathcal{H}
    \frac{d \boldsymbol{\Psi}}{d \tau}
    \right]
    =
    - \frac{1}{a^{2}}\nabla^{2}_{\boldsymbol{x}}\psi.
\end{align}

The Poisson equation now receives an additional contribution sourced by scalar-field fluctuations, which are coupled to the Klein--Gordon equation~(\ref{eq:mg_kg}). Since the Klein--Gordon equation is naturally written in Eulerian coordinates, we rewrite it in terms of Lagrangian coordinates by adding and subtracting the term $\nabla^{2}_{\boldsymbol{q}}\varphi$:
\begin{align}\label{eq:Psi_eq_mg}
    (J^{-1})_{ij}
    \nabla_{\boldsymbol{q}}\cdot
    \left[
    \hat{T}\boldsymbol{\Psi}(\boldsymbol{q},t)
    \right]
    =
    - \frac{\kappa}{2}\bar{\rho}\,
    \frac{1-J(\boldsymbol{q}, t)}{J(\boldsymbol{q}, t)}
    + \frac{1}{2}\nabla^{2}_{\boldsymbol{q}}\varphi
    + \frac{1}{2}
    \left[
    \nabla^{2}_{\boldsymbol{x}}\varphi
    -
    \nabla^{2}_{\boldsymbol{q}}\varphi
    \right](\boldsymbol{q},t),
\end{align}
where the final contribution is a geometrical correction arising from the transformation between Eulerian and Lagrangian coordinates. This term is commonly referred to as the \textit{frame-lagging} term~\cite{Aviles:2017aor}. 
%%%%%%%%%%%%%%%%%%%%%%%%%%%%%%%%%%%%%%%%%
%%%%%%%%%%%%%% FIGURE %%%%%%%%%%%%%%%%%%%
%%%%%%%%%%%%%%%%%%%%%%%%%%%%%%%%%%%%%%%%%
\begin{figure}[h]
    \centering
    \includegraphics[width=\linewidth]{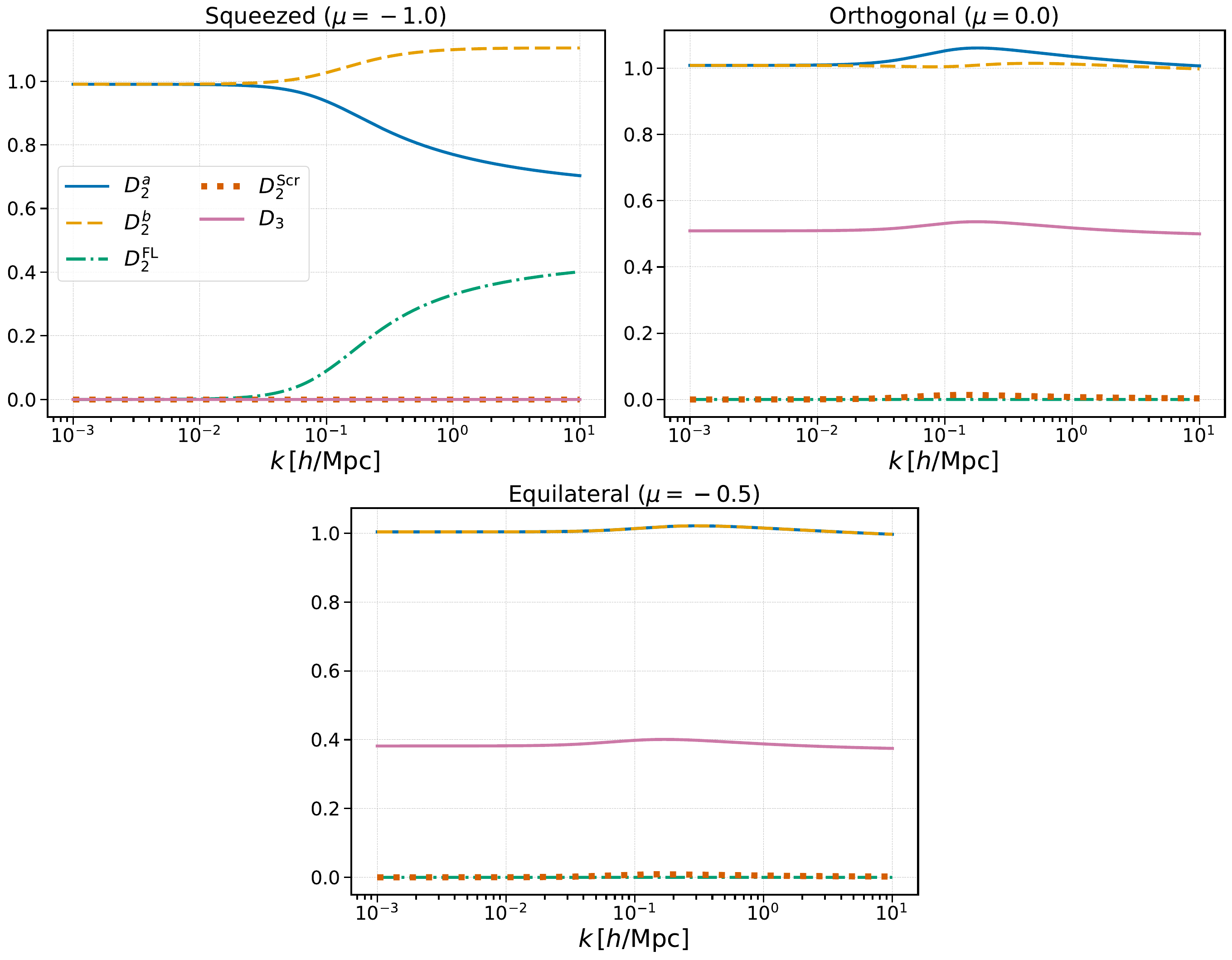}
    \caption{$f(R)$ gravity scale dependence of the normalized second- and third-order Lagrangian growth factors for different triangle configurations in Fourier space: squeezed ($\mu=-1$), orthogonal ($\mu=0$), and equilateral ($\mu=-0.5$). The quantities displayed are the second-order growth factors $D_{2}^{a}$, $D_{2}^{b}$, the frame-lagging contribution $D_{2}^{\rm FL}$, the screening contribution $D_{2}^{\rm Scr}$, and the third-order growth factor $D_{3}$. The strongest scale dependence appears in the squeezed configuration.}
    \label{fig:growth}
\end{figure}
%%%%%%%%%%%%%%%%%%%%%%%%%%%%%%%%%%%%%%%%%
%%%%%%%%%%%%%%%%%%%%%%%%%%%%%%%%%%%%%%%%%
Fourier transforming (\ref{eq:mg_kg}), where $\boldsymbol{k}$ now is with respect to $\boldsymbol{q}$, one gets:
\begin{align}
-\frac{k^2}{2a^2}\,\varphi(\boldsymbol{k}) 
&= -\left(A(k) - A_0\right)\tilde{\delta}(\boldsymbol{k})
+ \frac{k^2/a^2}{6\Pi(k)}\,\delta I(\boldsymbol{k}) \\
&\quad - \frac{(3 + 2\omega_{\rm BD})\,k^2/a^2}{3\Pi(k)}\,
\frac{1}{2a^2}
\left[\left(\nabla_x^2 \varphi - \nabla^2 \varphi\right)\right](\boldsymbol{k}) \, ,
\end{align}
with the following definitions:
\begin{align}
A(k) &= 4\pi G \bar{\rho} \left( 1 + \frac{k^2/a^2}{3\Pi(k)} \right), \\[6pt]
\Pi(k) &= \frac{1}{3a^2} \left( (3 + 2\omega_{\rm BD})\,k^2 + M_1 a^2 \right), \\[6pt]
A_0 &= A(k=0,t) = 4\pi G \bar{\rho}.
\end{align}
These equations are the same ones presented in~\cite{Aviles:2017aor}, and we have decided to follow their definition to make it easier to compare step by step with their previously found results. Therefore, we point the reader to the full equations of motion of $D^{(1)}(k)$, $D^{(2)}(\boldsymbol{k}_{1},\boldsymbol{k}_{2})$ and $D^{(3, \mathrm{symm.})}(\boldsymbol{k}_{1},\boldsymbol{k}_{2},\boldsymbol{k}_{3})$ in their paper, with the missing term of the third order screening having its full expression shown in~\cite{Aviles:2018qot}.

%%%%%%%%%%%%%%%%%%%%%%%%%%%%%%%%%%%%%%%%%
%%%%%%%%%%%%%% FIGURE %%%%%%%%%%%%%%%%%%%
%%%%%%%%%%%%%%%%%%%%%%%%%%%%%%%%%%%%%%%%%
\begin{figure}[h]
    \centering
    \includegraphics[width=\linewidth]{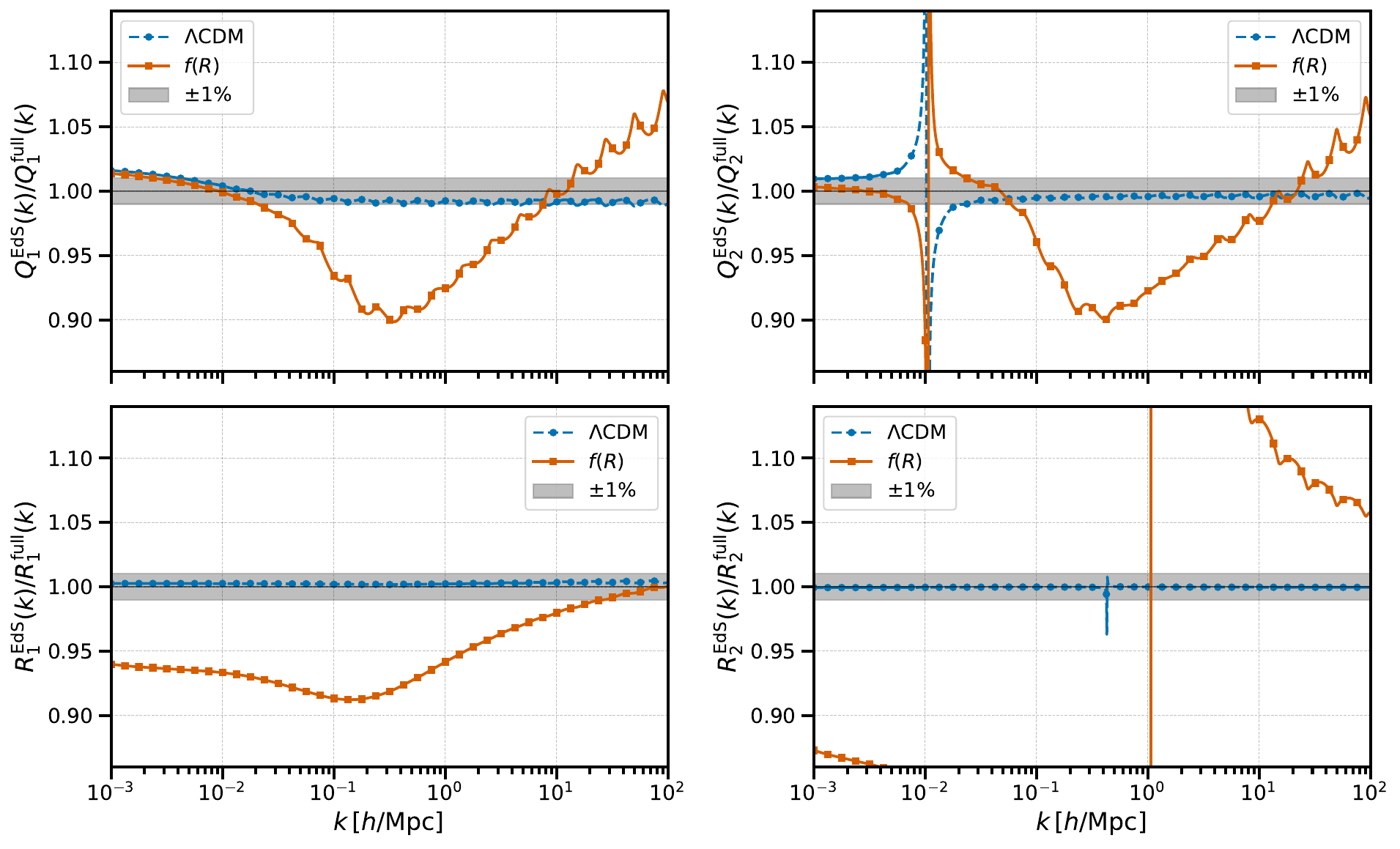}
    \caption{
    Ratios between the Einstein--de Sitter approximation and the exact time-dependent solutions for the loop functions $Q_{1}(k)$, $Q_{2}(k)$, $R_{1}(k)$, and $R_{2}(k)$ in both $\Lambda$CDM and $f(R)$ gravity. Dashed blue lines correspond to $\Lambda$CDM, while solid orange lines denote the $f(R)$ model. In the $\Lambda$CDM case, the EdS approximation reproduces the exact solutions with sub-percent accuracy over the full range of scales shown. In contrast, the modified-gravity case exhibits visible departures from unity, reflecting the failure of the EdS approximation in the presence of scale-dependent growth and screening effects. These deviations become more pronounced at mildly nonlinear scales, where mode coupling is strongest and the modified gravitational dynamics most strongly affect the higher-order kernels. Some of the wiggling in the curves are simply due to the numerical integration.}
    \label{fig:Q1_Q2_R1_Q3_ratios}
\end{figure}
%\begin{figure}[h]
%    \centering
%    \includegraphics[width=\linewidth]{Q1_Q2_R1_R2_EdS_over_full_ratios.pdf}
%    \caption{
%    $Q_{1}(k)$, $Q_{2}(k)$, $R_{1}(k)$, and $-R_{2}(k)$ functions computed for both $\Lambda$CDM and $f(R)$ gravity. Solid lines correspond to the exact time evolution obtained from the generalized growth functions, while markers show the Einstein--de Sitter (EdS) approximation for the corresponding cosmology. The modified-gravity model exhibits an overall enhancement of the amplitudes of the loop integrals relative to $\Lambda$CDM, reflecting the stronger clustering induced by the scale-dependent gravitational coupling. Deviations between the exact solutions and the EdS approximation become more pronounced in the $f(R)$ case, particularly around the peak scales where nonlinear mode coupling is strongest.}
%    \label{fig:Q1_Q2_R1_R2}
%\end{figure}
%%%%%%%%%%%%%%%%%%%%%%%%%%%%%%%%%%%%%%%%%
%%%%%%%%%%%%%%%%%%%%%%%%%%%%%%%%%%%%%%%%%

We have implemented a code that solves these equations of motion in $3.43$ seconds~\footnote{
The computation was performed on a local machine equipped with a 12th Gen Intel Core i5-12450HX CPU (8 cores, 12 threads), compiled with GCC 12.2.0. The code is written C++, parallelized with OpenMPI, and uses a Runge–Kutta (RK4) integrator from \textsc{boost} (\url{https://www.boost.org/}). Reported runtimes correspond to wall-clock time.} in a grid of $41\times41\times50$ ($|\boldsymbol{k_{1}}|$, $|\boldsymbol{k}_{2}|$, $\mu$), where $\mu=\boldsymbol{\hat{k}_{1}}\cdot\boldsymbol{\hat{k}}_{2}$. Even though the third order growth factor depends on three momenta, only the so-called double squeezed configuration of these quadrilaterals contribute, leaving only three degrees of freedom, making them possible to compute in the same loop as the second order growth factor.
We show the solutions of the growth factors in Figure~\ref{fig:growth}.

The strongest scale dependence in the modified gravity solutions appears in the squeezed configuration, consistent with the enhanced mode coupling typically exhibited by LPT/SPT kernels in the squeezed limit, where long and short wavelength modes interact most efficiently. From the growth functions, we can compute the full Lagrangian kernels,
Eq.~(\ref{eq:lag_kernels_gen}), and use them to evaluate the scalar
functions defined in Eqs.~(\ref{eq:Q1_lag})--(\ref{eq:R2_lag}), which are
constructed from the Lagrangian correlators. In Fig.~\ref{fig:Q1_Q2_R1_Q3_ratios}, we compare the full-kernel solutions for $f(R)$ and $\Lambda$CDM with their corresponding EdS approximations. As can be seen, the EdS approximation in modified gravity fails to capture the full scale-dependent physics encoded in the growth functions, making it a poor
approximation in this case. These $k$-dependent functions can then be Fourier transformed to Lagrangian coordinates, $\boldsymbol{q}$, from
which we extract the quantities defined in Eqs.~(\ref{eq:X})--(\ref{eq:V}).

%%%%%%%%%%%%%% FIGURE %%%%%%%%%%%%%%%%%%%
%%%%%%%%%%%%%%%%%%%%%%%%%%%%%%%%%%%%%%%%%
\begin{figure}[h]
    \centering
    \includegraphics[width=\linewidth]{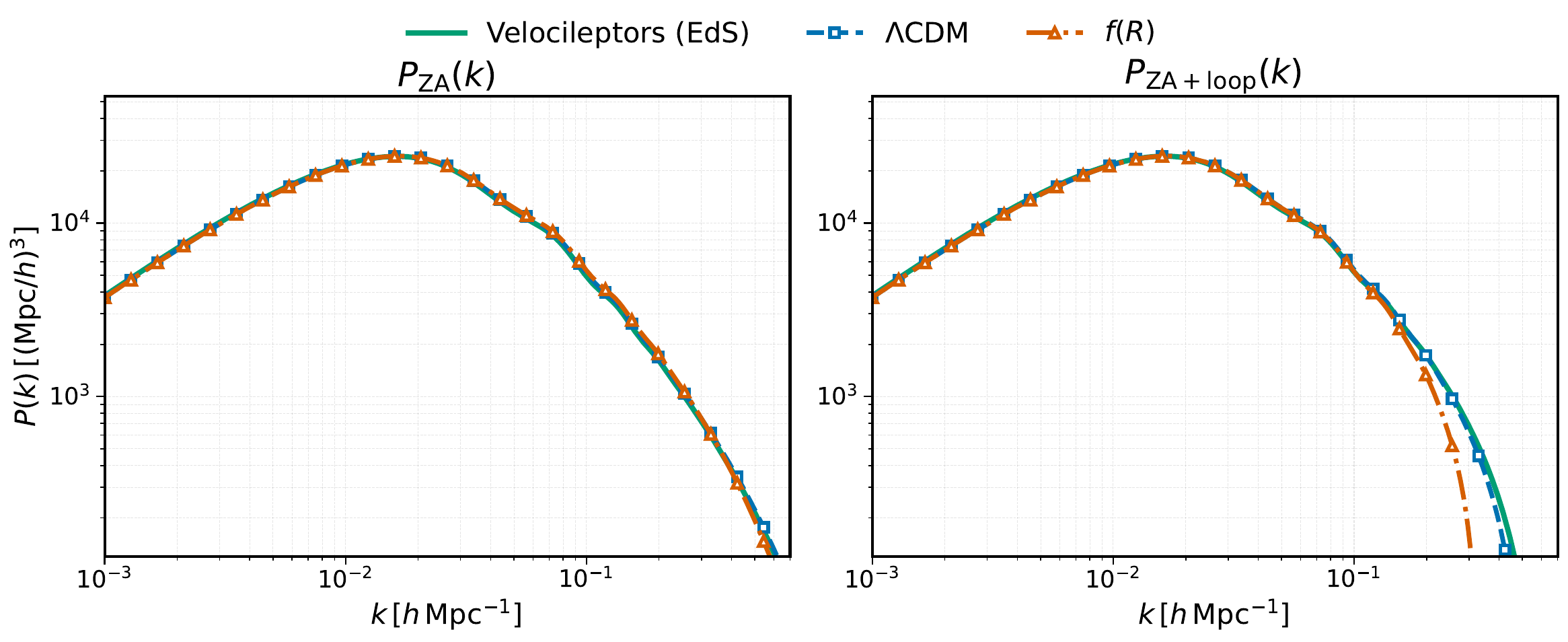}
    \caption{Matter power spectrum computed in LPT with and without loop
    corrections. We compare the EdS result obtained with
    \texttt{velocileptors} with the $\Lambda$CDM and $f(R)$ full-kernel
    predictions developed in this work. The exponential suppresion present in all LPT predictions is more accetuanted in the case of $f(R)$ as we have $\sigma^{2}_{L, f(R)}>\sigma^{2}_{L, \Lambda\mathrm{CDM}}$}
    \label{fig:pk_za_loop}
\end{figure}
%%%%%%%%%%%%%%%%%%%%%%%%%%%%%%%%%%%%%%%%%
%%%%%%%%%%%%%%%%%%%%%%%%%%%%%%%%%%%%%%%%%

To conclude this section, we show in Fig.~\ref{fig:pk_za_loop} the matter
power spectrum computed in LPT with and without loop corrections. We compare
the EdS prediction obtained with \texttt{velocileptors} against the $\Lambda$CDM and $f(R)$ full-kernel results computed in this work. As
discussed above, keeping only the linear displacement correlator $\boldsymbol{A}_{\rm linear}$ exponentiated corresponds to the fully non-perturbative Zel’dovich contribution. The inclusion of the remaining loop contributions then gives the one-loop LPT correction to this result.

\section{Biased tracers}\label{sec:biased}

The same LPT-based approach can be used to compute the one-loop corrected matter power spectrum for biased tracers. As shown in Ref.~\cite{Matsu2}, the starting point is the analogue of mass conservation for tracers:
\begin{equation}\label{eq:tracer_conser}
    \left[1+\delta_{X}(\boldsymbol{x},t)\right]d^3\boldsymbol{x}
    =
    \left[1+\delta_{X}(\boldsymbol{q},t_{0})\right]d^3\boldsymbol{q},
\end{equation}
where $\delta_{X}$ denotes the overdensity field of a given tracer population. Contrary to the matter case, however, Eq.~(\ref{eq:tracer_conser}) should not be interpreted as an exact conservation law, since the number density of tracers is generally not conserved during the evolution of the Universe. For example, dark matter haloes undergo mergers and other nonlinear processes that modify their abundance over time.

Nevertheless, within the framework of a local Lagrangian bias model, the term on the right-hand side of Eq.~(\ref{eq:tracer_conser}) is written as~\cite{White:2014gfa,LEFT,Aviles:2018thp,Matsu2}
\begin{equation}
    1 + \delta_{X}(\boldsymbol{q},t_{0})
    =
    F[\delta_{\rm L}(\boldsymbol{q})],
\end{equation}
where $\delta_{\rm L}(\boldsymbol{q})$ is the linear Gaussian initial matter density field. One then obtains
\begin{align}
    1+\delta_{X}(\boldsymbol{x},t)
    &=
    \left|
    \frac{\partial \boldsymbol{x}}
    {\partial\boldsymbol{q}}
    \right|^{-1}
    F[\delta_{\rm L}(\boldsymbol{q})]
    \nonumber\\
    &=
    \int d^{3}q~
    F[\delta_{\rm L}(\boldsymbol{q})]
    \delta^{\rm D}
    \left(
    \boldsymbol{x}
    -\boldsymbol{q}
    -\boldsymbol{\Psi}(\boldsymbol{q},t)
    \right),
\end{align}
where $\boldsymbol{\Psi}(\boldsymbol{q},t)$ is the Lagrangian displacement field.

The general expression for the LPT power spectrum of biased tracers is then given by
\begin{equation}\label{eq:pk_biased_lpt}
    (2\pi)^{3}\delta^{\rm D}(\boldsymbol{k})
    + P_{\rm LPT}(k)
    =
    \int d^{3}q~
    e^{i\boldsymbol{k}\cdot\boldsymbol{q}}
    \int
    \frac{d\lambda_{1}}{2\pi}
    \frac{d\lambda_{2}}{2\pi}
    ~
    \tilde{F}(\lambda_{1})
    \tilde{F}(\lambda_{2})
    \left\langle
    e^{i\left[
    \lambda_{1}\delta_{\rm L}(\boldsymbol{q}_{1})
    +
    \lambda_{2}\delta_{\rm L}(\boldsymbol{q}_{2})
    \right]
    -i\boldsymbol{k}\cdot
    \left[
    \boldsymbol{\Psi}(\boldsymbol{q}_{1})
    -
    \boldsymbol{\Psi}(\boldsymbol{q}_{2})
    \right]}
    \right\rangle,
\end{equation}
where $\tilde{F}(\lambda_i)$ denotes the Fourier transform of the functional
$F[\delta_{\rm L}(\boldsymbol{q}_{i})]$, and
$\boldsymbol{q}=\boldsymbol{q}_{1}-\boldsymbol{q}_{2}$. In this framework, Eq.~(\ref{eq:pk_biased_lpt}) should be interpreted as the power spectrum of discrete tracers whose statistical properties are determined by fields constructed from the initial density field $\delta_{\rm L}(\boldsymbol{q})$. The bias expansion is written as
\begin{align}\label{eq:F_delta_q}
    F[\delta_{\rm L}(\boldsymbol{q})]
    &=
    1+\delta_{X}(\boldsymbol{q},t_{0})
    \nonumber\\
    &=
    1
    +
    b_{1}\delta_{\rm L}(\boldsymbol{q})
    +
    b_{2}
    \left(
    \delta^{2}_{\rm L}(\boldsymbol{q})
    -
    \langle\delta^{2}_{\rm L}(\boldsymbol{q})\rangle
    \right)
    \nonumber\\
    &\quad
    +
    b_{s^2}
    \left(
    s^{2}(\boldsymbol{q})
    -
    \langle s^{2}\rangle
    \right)
    +
    b_{3}\mathcal{O}_{3}(\boldsymbol{q})
    + \dots,
\end{align}
where
\begin{equation}
    s^{2}(\boldsymbol{q})
    =
    s_{ij}(\boldsymbol{q})s_{ij}(\boldsymbol{q}),
\end{equation}
and
\begin{equation}
    s_{ij}(\boldsymbol{q})
    =
    \left(
    \frac{\partial_{i}\partial_{j}}{\partial^{2}}
    -
    \frac{\delta_{ij}}{3}
    \right)
    \delta_{\rm L}(\boldsymbol{q}),
\end{equation}
is the tidal shear tensor, and the derivatives are taken with respect to the Lagrangian coordinates, $\boldsymbol{q}$.

In this work, we consider the bias expansion only up to the tidal shear operator. However, many other works and numerical implementations also include third-order bias operators and curvature bias terms~\cite{LSSTDarkEnergyScience:2023qfp,Kokron:2021faa,deBelsunce:2025gci,Hadzhiyska:2021xbv,Banerjee:2021cmi,Rubiola:2025ntk,Pellejero-Ibanez:2022efv,Garcia-Garcia:2024gzy}.
From (\ref{eq:F_delta_q}) we can see that now we will be interested in dealing with Lagrangian correlators that are not only functions of the displacement field, but also of the initial density field, in this way~\cite{CLEFT,ChenPhD} introduced a more general way of writing these correlators:
\begin{equation}
    C_{i_{1}i_{2}\dots}^{n_{1}n_{2}(m_{1}m_{2}\dots)} = \langle \delta_{1}^{n_{1}}\delta_{2}^{n_{2}}\Delta\boldsymbol{\Psi}^{(m_{1})}_{i_{1}}\Delta\boldsymbol{\Psi}^{(m_{2})}_{i_{2}}\dots \rangle,
\end{equation}
where the term in parentheses are the order of the Lagrangian solution, and the dots are further powers of the difference between displacement fields, note also that quantities with parentheses omitted imply the sums to all orders, $C^{00}_{ij} = C^{00(11)}_{ij} + C^{00(22)}_{ij}+C^{00(13)}_{ij}+C^{00(31)}_{ij} + \dots$, up to quadratic order.

Similarly to the one-loop corrected matter power spectrum case,
Appendix B of Ref.~\cite{CLEFT} shows that the computation of the
Lagrangian correlators can ultimately be reduced to the evaluation of a
set of scalar functions in $\boldsymbol{q}$-space. Since we are working
within a modified theory of gravity, however, the $Q_i$ and $R_i$
functions derived in their Appendix B cannot be directly applied.
Instead, we evaluate the corresponding integrals following the expressions
presented in Refs.~\cite{AA2,Valogiannis:2019xed}\footnote{Specifically, the $Q_{i}(k)$ and $R_{i}(k)$ implemented in Appendix D of~\cite{AA2}.}, which we have also
independently verified in this work.\footnote{The $Q_i(k)$ and $R_i(k)$
functions are evaluated using trapezoidal radial integration and
Gauss--Legendre angular quadrature on a $1750 \times 512$ grid.}

\begin{figure}[h]
    \centering
    \includegraphics[width=\linewidth]{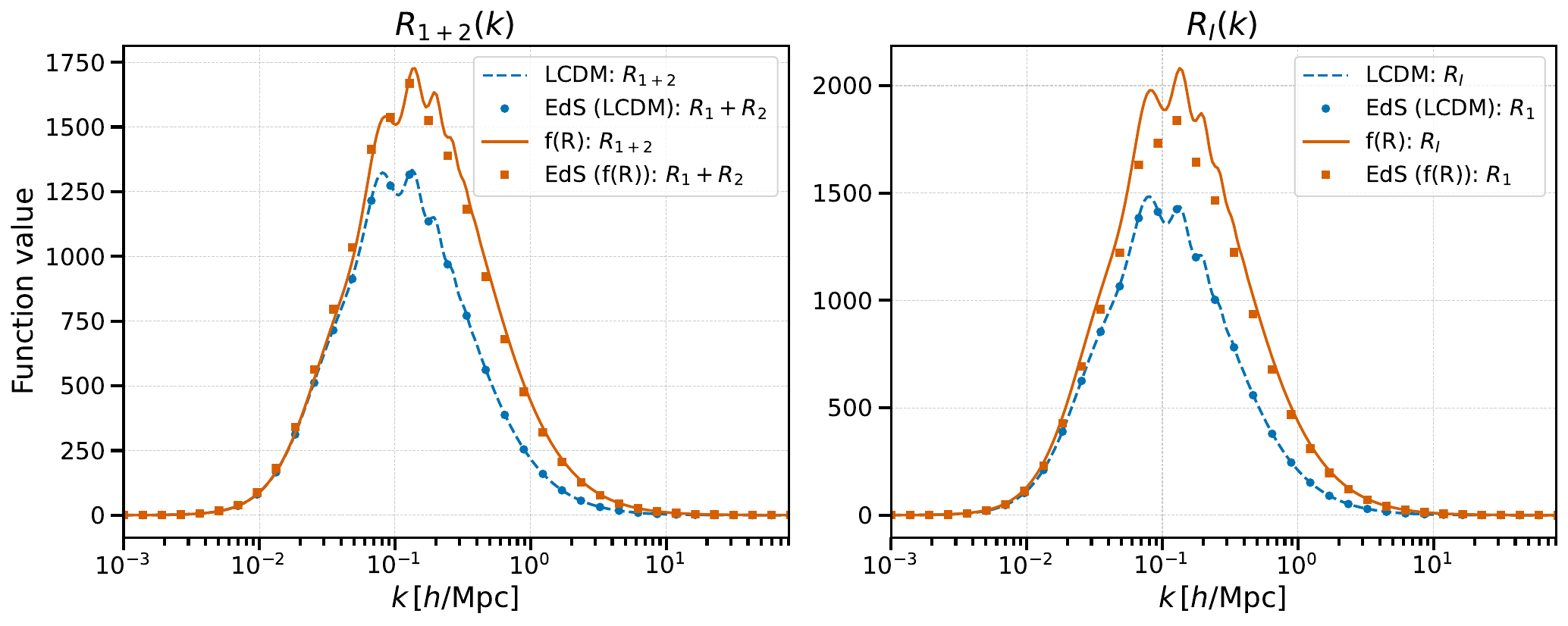}
    \caption{Comparison between the exact one-loop Lagrangian perturbation theory integrals $R_{1+2}(k) \equiv R_{1}(k)+R_{2}(k)$ and $R_{I}(k)$, and their Einstein--de Sitter (EdS) approximations, computed for both $\Lambda$CDM and $f(R)$ gravity. Solid lines correspond to the full generalized calculation, while markers denote the EdS approximation. For the $\Lambda$CDM case, the EdS approximation reproduces the exact result remarkably well across the full $k$ range considered. In contrast, noticeable deviations appear in the $f(R)$ model, particularly around the mildly nonlinear regime where the loop integrals peak, indicating that the EdS relations become less accurate in the presence of scale-dependent modified-gravity effects.
    }
    \label{fig:R_panel_comparison}
\end{figure}

As shown in Refs.~\cite{AA2,Valogiannis:2019xed}, generalized
cosmologies introduce additional $k$-dependent functions which, in the EdS limit, reduce to simple combinations of the standard EdS quantities, namely
\begin{equation}
R_I^{\mathrm{EdS}} = R_1^{\mathrm{EdS}}, \qquad
Q_I^{\mathrm{EdS}} = Q_1^{\mathrm{EdS}}, \qquad
R_{1+2}^{\mathrm{EdS}}
=
R_1^{\mathrm{EdS}}
+
R_2^{\mathrm{EdS}} .
\end{equation}
To better understand the impact of using and not using the EdS Lagrangian kernels in $\Lambda$CDM and $f(R)$, we show in
Fig.~\ref{fig:R_panel_comparison} the functions $R_{1+2}(k)$ and $R_I(k)$ computed from Eq.~(C.24) of Ref.~\cite{AA2}, and compare them against their corresponding EdS combinations,
$R_1(k)+R_2(k)$ and $R_1(k)$, respectively.
%%%%%%%%%%%%%%%%%%%%%%%%%%%%%%%%%%%%%%%%%
%%%%%%%%%%%%%% FIGURE %%%%%%%%%%%%%%%%%%%
%%%%%%%%%%%%%%%%%%%%%%%%%%%%%%%%%%%%%%%%%
%%%%%%%%%%%%%%%%%%%%%%%%%%%%%%%%%%%%%%%%%
%%%%%%%%%%%%%%%%%%%%%%%%%%%%%%%%%%%%%%%%%
Once again, we find that the EdS approximation provides an excellent description for the $\Lambda$CDM case, as indicated by the near-perfect overlap between the blue square markers and the dashed blue line. However, the same does not hold for $f(R)$ gravity, where the dark-orange square markers exhibit significant deviations from the full-kernel computation, particularly in the mildly nonlinear regime, $k \sim 0.1$--$0.4\,h\,\mathrm{Mpc}^{-1}$.

\begin{figure}[h]
    \centering
    \includegraphics[width=\linewidth]{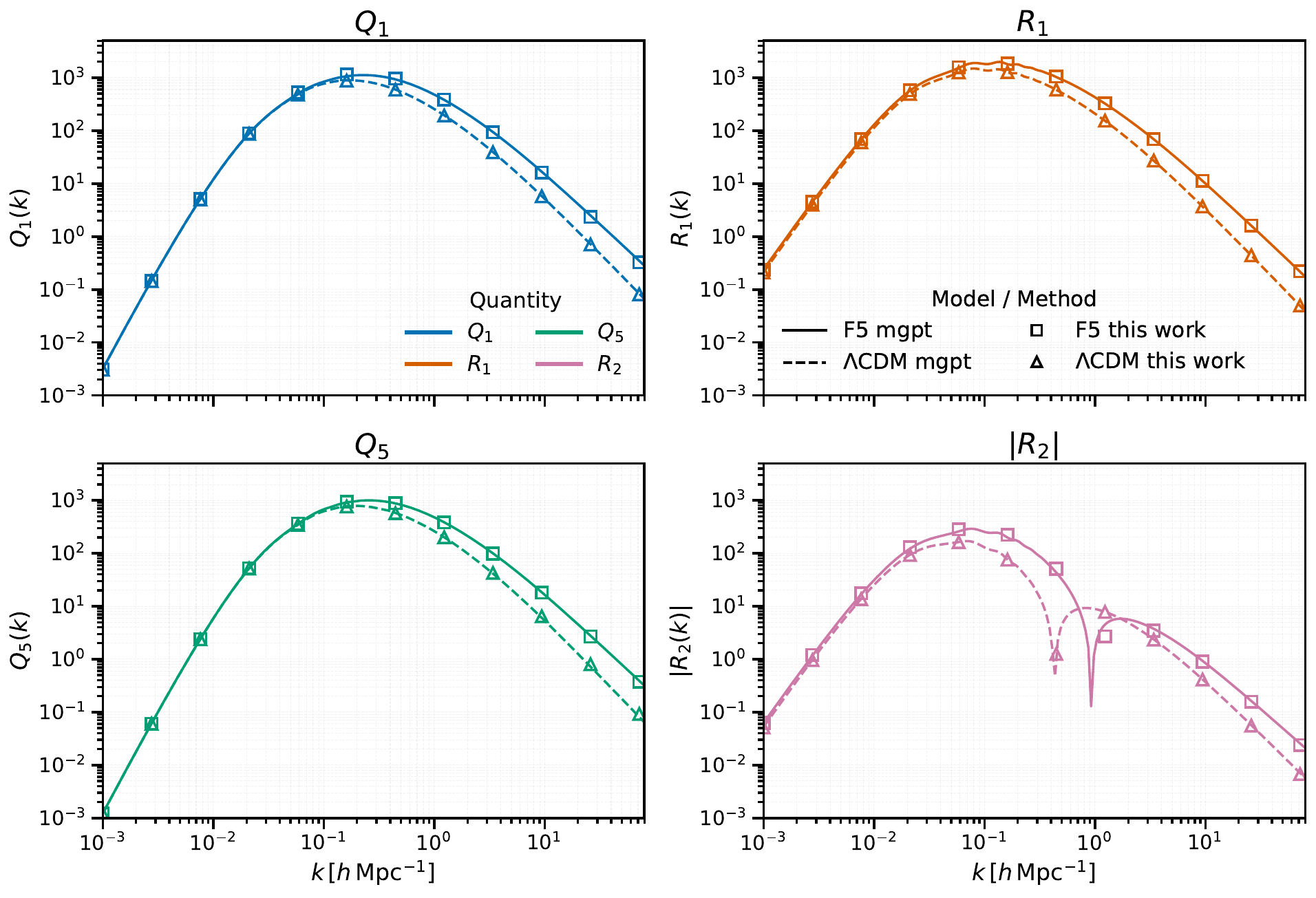}
    \caption{Comparison between integrals computed in this work and the predictions from the \texttt{mgpt} code for both $\Lambda$CDM and $f(R)$ gravity. The panels show the quantities $Q_{1}(k)$, $R_{1}(k)$, $Q_{5}(k)$, and $\left|R_{2}(k)\right|$ as functions of wavenumber $k$. Solid and dashed lines correspond to the \texttt{mgpt} predictions for the $f(R)$ and $\Lambda$CDM models, respectively, while square and triangular markers denote the results obtained in this work. Overall, excellent agreement is observed over the full range of scales considered, validating the numerical implementation of the generalized LPT kernels and loop integrals in modified gravity cosmologies in this work.
    }
    \label{fig:Q1_R1_Q5_R2_mgpt_gui}
\end{figure}

In order to further validate the robustness of our implementation, we compare our results against the \texttt{mgpt} code\footnote{\url{https://github.com/alejandroaviles/mgpt.git}}, presented in Ref.~\cite{AA2}, which is also capable of computing the same integrals in both $k$- and $q$-space employed throughout this work. In Fig.~\ref{fig:Q1_R1_Q5_R2_mgpt_gui}, we show the comparison for four representative functions in both $\Lambda$CDM and the F5 model.\footnote{The F5 model corresponds to the parameter choice $\log_{10}|\bar{f}_{R}|=-5$.}
%%%%%%%%%%%%%%%%%%%%%%%%%%%%%%%%%%%%%%%%%
%%%%%%%%%%%%%% FIGURE %%%%%%%%%%%%%%%%%%%
%%%%%%%%%%%%%%%%%%%%%%%%%%%%%%%%%%%%%%%%%
%%%%%%%%%%%%%%%%%%%%%%%%%%%%%%%%%%%%%%%%%
%%%%%%%%%%%%%%%%%%%%%%%%%%%%%%%%%%%%%%%%%
As shown in Fig.~\ref{fig:Q1_R1_Q5_R2_mgpt_gui}, our results are in excellent agreement with \texttt{mgpt}, providing an important consistency check of our implementation and numerical pipeline. 

As in the one-loop matter power spectrum case, (\ref{eq:pk_lpt_cdm}), the real-space biased-tracer power spectrum is computed from the Fourier transform of these $k$-space scalar functions. We then write here the definition of the biased-tracer power spectrum as introduced in Appendix B of Ref.~\cite{Vlah:2016bcl}:
\begin{align}\label{eq:pk_1loop_biased}
P(k)&= \int d^{3}\mathbf{q}\;
e^{i k \mu q}
e^{-\frac{1}{2}k_i k_j A^{\rm lin}_{ij}}
\Bigg\{
1
-\frac{1}{2}k^{2}\left(X^{\rm loop}+Y^{\rm loop}\mu^{2}\right)
-\frac{i}{6}k^{3}\left(\tilde V\,\mu+\tilde T\,\mu^{3}\right)
\nonumber\\[0.3em]
&+ b_{1}\left(
2ikU\mu
-k^{2}\left(X^{10}+Y^{10}\mu^{2}\right)
\right)
+ b_{1}^{2}\left(
\xi_{\rm lin}^{2}
-k^{2}U_{\rm lin}^{2}\mu^{2}
+ikU^{11}\mu
\right)
\nonumber\\[0.3em]
&+ b_{2}\left(
-k^{2}U_{\rm lin}^{2}\mu^{2}
+ikU^{20}\mu
\right)
+2i b_{1}b_{2}k\xi_{\rm lin}U_{\rm lin}\mu
+\frac{b_{2}^{2}}{2}\xi_{\rm lin}^{2}
\nonumber\\[0.3em]
&+ b_{s}\left(
-k^{2}(X_{s}+Y_{s}\mu^{2})
+2ikV^{10}\mu
\right)
+2i b_{1}b_{s}kV^{12}\mu
+b_{2}b_{s}\chi
+b_{s}^{2}\zeta
\Bigg\}.
\end{align}
All functions appearing in the integrand are scalar functions of $q$, and
their definitions can be found in
Refs.~\cite{CLEFT,ChenPhD,Vlah:2016bcl} in EdS, and in~\cite{AA2,Valogiannis:2019xed} for the modified gravity counterpart. To compute these quantities, we
follow the same strategy implemented in \texttt{velocileptors} and described in several references~\cite{Schmittfull:2016jsw,
Schmittfull:2016yqx,Fang:2016wcf,Simonovic:2017mhp,Tomlinson:2020xbf}, based on Hankel transforms.

Inspecting the integral form of the $q$-space functions entering Eq.~(\ref{eq:pk_1loop_biased}), these terms can be cast into a generic form:
\begin{align}
\int d^{3}q\;
&e^{ikq\mu-\frac{1}{2}k_i k_j A_{ij}^{\rm lin}}\,
\mu^{m}\,f(q)
=\nonumber\\
&\sum_{n=0}^{\infty} 4\pi
\int dq\, q^{2}\,
e^{-\frac{1}{2}k^{2}\left(X^{\rm lin}+Y^{\rm lin}\right)}
f_{n}^{m}\!\left(k^{2}Y_{\rm lin}\right)
\left(\frac{kY_{\rm lin}}{q}\right)^{n}
f(q)\,
j_{n}(kq),
\end{align}
where the angular dependence has been reorganized into a sum over spherical Bessel functions $j_n(kq)$. This representation makes the remaining radial integrals suitable for fast evaluation through Hankel transforms.
%%%%%%%%%%%%%%%%%%%%%%%%%%%%%%%%%%%%%%%%%
%%%%%%%%%%%%%% FIGURE %%%%%%%%%%%%%%%%%%%
%%%%%%%%%%%%%%%%%%%%%%%%%%%%%%%%%%%%%%%%%
\begin{figure}[h]
    \centering
    \includegraphics[width=\linewidth]{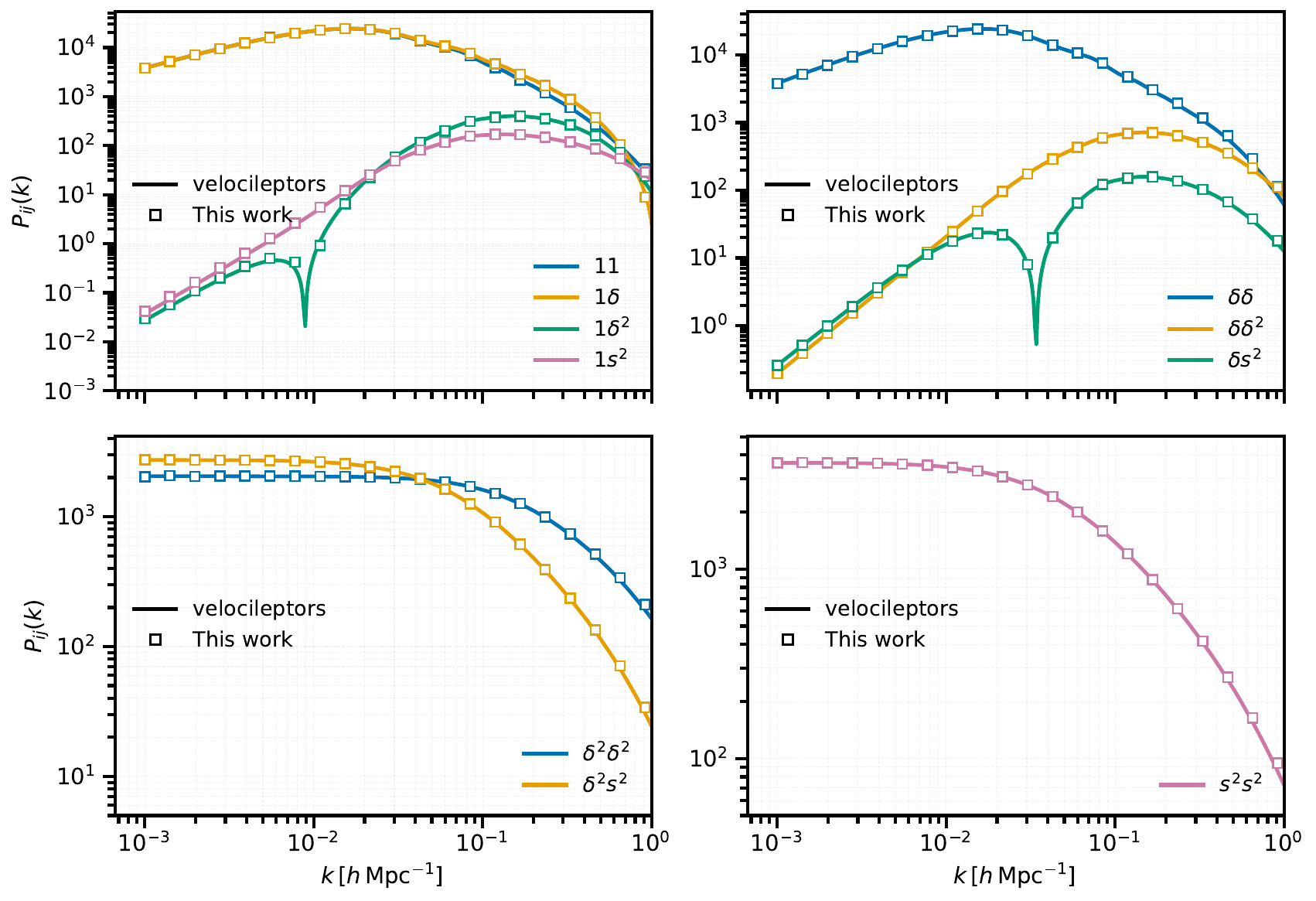}
    \caption{Comparison between the one-loop LPT operator spectra computed in this work and the predictions from the \texttt{velocileptors} implementation for $\Lambda$CDM using the EdS approximation. Solid lines denote the \texttt{velocileptors} results, while square markers correspond to the calculations performed in this work. The panels display the auto- and cross-spectra associated with the HEFT operators $\delta$, $\delta^{2}$, and $s^{2}$. Excellent agreement is observed across the full range of scales considered, validating the implementation of the generalized Lagrangian kernels and the numerical evaluation of the loop contributions.
    }
    \label{fig:gui_vel_P_ij}
\end{figure}
%%%%%%%%%%%%%%%%%%%%%%%%%%%%%%%%%%%%%%%%%
%%%%%%%%%%%%%%%%%%%%%%%%%%%%%%%%%%%%%%%%%

In practice, we apply this procedure term by term to all contributions entering Eq.~(\ref{eq:pk_1loop_biased}). In the EdS case, part of this structure can be pretabulated in terms of fixed kernels. In the present $f(R)$ case, however, these kernels are replaced by scale- and time-dependent functions, implying that the $k$-dependent coefficients entering the Hankel integrals must be recomputed for each model evaluation. Nevertheless, the FFTLog infrastructure itself, including the logarithmic $q$-grid and the Hankel-transform setup, can still be reused,
keeping the calculation computationally efficient.

Using FFTLog-based integration methods, the computation of the
$q$-space functions requires only approximately $0.17\,\mathrm{s}$, while the full evaluation of Eq.~(\ref{eq:pk_1loop_biased}) takes $10.8\,\mathrm{s}$. In this work we use $n_{\rm max}=5$, which already provides excellent agreement with \texttt{velocileptors}, at the
level of better than $2\%$ up to $k=0.3\,h\,\mathrm{Mpc}^{-1}$, and the differences between the two at high $k$ is seen due to the heavy exponential suppression. We show a comparison between the two codes in Fig.~\ref{fig:gui_vel_P_ij}, noting that we are comparing our $\Lambda$CDM kernels with their EdS kernels, as we have already stated that the level of agreement between the two are excellent. The direct comparison between the code developed in this work with the estabilished \texttt{velocileptors} one provides a strong argument for the consistency of our pipeline.

%%%%%%%%%%%%%%%%%%%%%%%%%%%%%%%%%%%%%%%%%
%%%%%%%%%%%%%% FIGURE %%%%%%%%%%%%%%%%%%%
%%%%%%%%%%%%%%%%%%%%%%%%%%%%%%%%%%%%%%%%%
\begin{figure}[h]
    \centering
    \includegraphics[width=\linewidth]{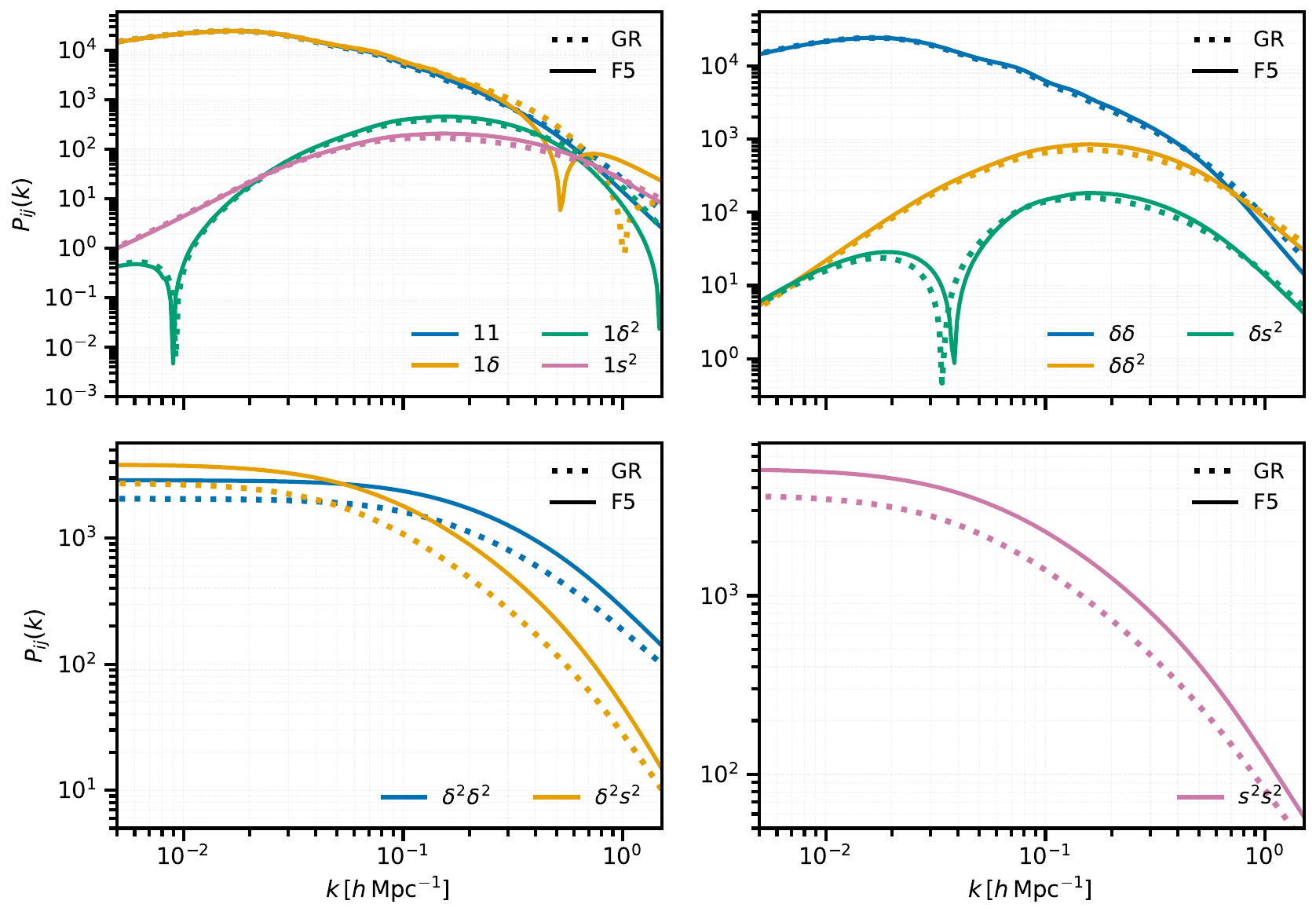}
    \caption{Same as Figure~\ref{fig:gui_vel_P_ij} but showing $\Lambda$CDM and F5.
    }
    \label{fig:gui_GR_F5_P_ij}
\end{figure}
%%%%%%%%%%%%%%%%%%%%%%%%%%%%%%%%%%%%%%%%%
%%%%%%%%%%%%%%%%%%%%%%%%%%%%%%%%%%%%%%%%%

We conclude this section by presenting the CLPT basis power spectra in GR and in the F5 modified gravity model. The results largely follow the expected behaviour, especially for basis spectra involving either the usual matter power spectrum and the one from the linear density field, denoted by the subscript ``1'' and the $\delta_{\rm L}(\boldsymbol{q})$ operator, respectively. In these cases, modified gravity generally enhances clustering on small scales relative to the GR prediction. In contrast, spectra involving the $\delta^2_{\rm L}(\boldsymbol{q})$ and $s^2(\boldsymbol{q})$ Lagrangian fields exhibit a more intricate scale dependence, as these operators are more sensitive to the coupling between long- and short-wavelength modes. This leads to a stronger enhancement in their amplitudes relative to GR, as illustrated in the lower panels of Fig.~\ref{fig:auto_F5_pred} and Fig.~\ref{fig:cross_F5_pred}.

One final comment concerns the usual addition of counterterms that promote the CLPT approach into an EFT description, an issue discussed in~\cite{Rodriguez-Meza:2023rga,Euclid:2023bgs} in the Eulerian framework. In particular, Ref.~\cite{Rodriguez-Meza:2023rga} argues that the screening contributions appearing in the Lagrangian kernels are highly degenerate with the usual EFT counterterms. For the case of $f(R)$ gravity, which naturally exhibits chameleon screening, the impact of these screening corrections on the power spectrum multipoles can be effectively absorbed by counterterms of the standard EFT form, proportional to $k^{2}P_{\rm L}(k)$. This suggests that, rather than explicitly including screening contributions in the perturbative kernels, their net effect can be accounted for through a redefinition of the EFT coefficients. In that sense, their analysis motivates neglecting explicit screening terms and absorbing their contribution into the EFT counterterms.

Therefore, the inclusion of the standard EFT counterterms may ultimately provide a computationally efficient alternative to the explicit treatment of screening corrections in the kernels presented in this work, since the evaluation of the third-order screening contributions entering $D^{(3,\mathrm{symm.})}$ constitutes one of the main numerical bottlenecks in our implementation. We further speculate that, for theories in which Vainshtein screening is the dominant mechanism, a similar degeneracy may arise involving the shear-bias operators and the additional higher-derivative curvature terms that characterize these theories and suppress the fifth force in dense environments. Exploring this possibility in detail is beyond the scope of the present work and is left for future investigation.

\section{Hybrid Effective Field Theory}\label{sec:heft}

Having presented the analytical derivation and numerical implementation
of the LPT framework for computing the matter and biased-tracer power
spectra, we now turn to the Hybrid Effective Field Theory (HEFT)
approach and our simulation results. As first proposed
in Ref.~\cite{Modi:2019qbt}, there exists a remarkable correspondence
between the functional approach to biased tracers in LPT introduced
in Ref.~\cite{Matsu2} and the evolution of dark matter particles in
\textit{N}-body simulations.

In the analytical formalism, the smoothed initial density field,
$\delta_{\rm L}$, is obtained from standard Einstein--Boltzmann solvers.
This same field also serves as the starting point for generating the
initial conditions of \textit{N}-body simulations, which are now capable
of describing the clustering of dark matter on small scales with high
precision. 
%%%%%%%%%%%%%%%%%%%%%%%%%%%%%%%%%%%%%%%%%
%%%%%%%%%%%%%% FIGURE %%%%%%%%%%%%%%%%%%%
%%%%%%%%%%%%%%%%%%%%%%%%%%%%%%%%%%%%%%%%%
\begin{figure}[h]
    \centering
    \includegraphics[width=\linewidth]{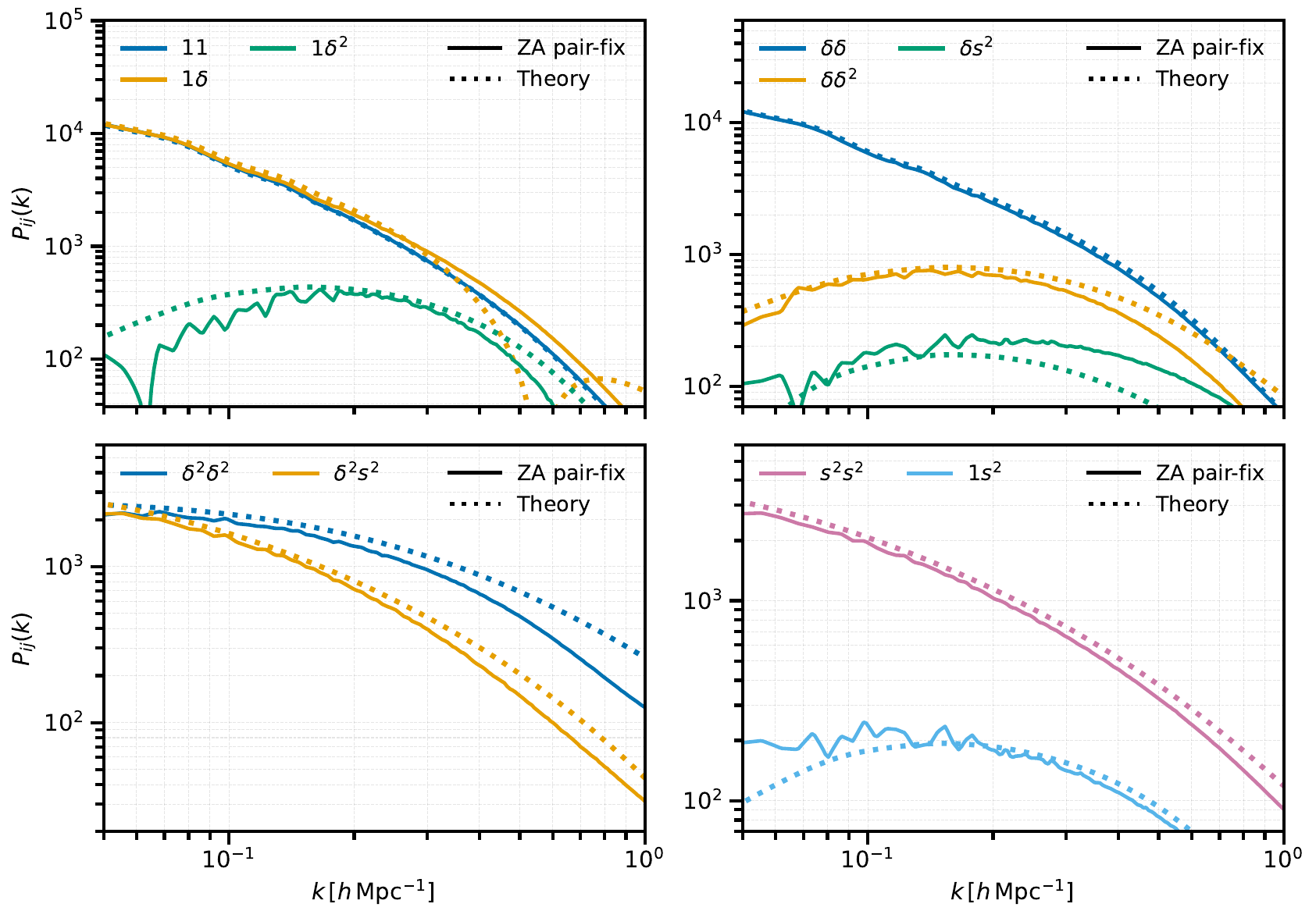}
    \caption{Comparison between the HEFT operator spectra measured from Zel'dovich approximation (ZA) particle snapshots and the corresponding theoretical predictions obtained from the generalized Lagrangian framework in $f(R)$ gravity. The ZA measurements are constructed from particle grids evolved according to the Zel'dovich approximation using pair-fixed initial conditions. Solid lines denote the measured ZA spectra, while dotted lines show the theoretical calculations. The panels display the auto- and cross-spectra associated with the HEFT operators $\delta$, $\delta^{2}$, and $s^{2}$. Overall, the theoretical predictions reproduce the broadband behaviour and relative amplitudes of the measured ZA spectra over the full range of scales considered, with the largest deviations appearing in spectra involving higher-order operators at nonlinear scales.
    }
    \label{fig:f5_th_za_P_ij}
\end{figure}
%%%%%%%%%%%%%%%%%%%%%%%%%%%%%%%%%%%%%%%%%
%%%%%%%%%%%%%%%%%%%%%%%%%%%%%%%%%%%%%%%%%
The generation of simulation initial conditions begins by
constructing a three-dimensional Gaussian random field,
$\delta_{\rm L}(\boldsymbol{q})$, sampled on a regular Lagrangian grid. In
this setup, each Lagrangian coordinate $\boldsymbol{q}$ is uniquely
associated with a simulation particle through its particle ID.

The standard procedure for estimating the matter power spectrum from a
simulation consists of depositing particles onto a fixed grid using a
mass-assignment scheme and subsequently Fourier transforming the density
field using FFTs. In the matter case, each particle contributes equally,
corresponding to a uniform weighting. In the functional-bias approach,
however, each particle is instead weighted by the value of a given
Lagrangian operator evaluated at its initial position, such as
$\delta_{\rm L}(\boldsymbol{q})$,
$\delta_{\rm L}^{2}(\boldsymbol{q})$, or
$s^{2}(\boldsymbol{q})$. Following this procedure, one can estimate the
basis spectra associated with the theoretical predictions derived in the
previous section,
\begin{equation}
    P_{ij}(k)
    =
    \sum_{i,j=\delta_{\rm L},\,\delta_{\rm L}^{2},\,s^{2}}
    b_{i}b_{j}\,
    \mathcal{O}_{i}\mathcal{O}_{j}.
\end{equation}

While this procedure has already been extensively explored for
$\Lambda$CDM, $w$CDM, and $w_{0}w_{a}$CDM cosmologies, leading to the
construction of several emulators~\cite{DeRose:2018xdj,Zennaro:2021bwy,
Zhou:2025iiu}, here we present the corresponding basis-spectrum
computations for F5 gravity and compare them against their theoretical
predictions. Furthermore, it is also instructive to compute the same
basis spectra from a Zel'dovich realization, \textit{i.e.}, a particle
snapshot generated using the Zel'dovich approximation.

We compare our theoretical predictions obtained with the code
developed in this work against the spectra measured from the Zel'dovich
particle realization in Fig.~\ref{fig:f5_th_za_P_ij}. To reduce sample
variance, we generated Zel'dovich realizations using paired-fixed
initial conditions. We find good agreement between the theoretical predictions and the grid-based measurements. The agreement is generally better for the auto-spectra, $P_{ii}(k)$, which is expected since auto-power spectra are positive-definite quantities and therefore less sensitive to fractional fluctuations induced by sample variance and finite-volume effects. In contrast, cross-spectra are not positive-definite and can exhibit larger relative deviations, particularly on large scales where cancellations and noise become more important. This effect is especially pronounced for composite Lagrangian operators such as $\delta_{\rm L}^{2}(\boldsymbol{q})$ and $s^{2}(\boldsymbol{q})$, whose cross-correlations can be partially suppressed by renormalization or subtraction terms, resulting in small net signals and consequently larger apparent discrepancies in relative comparisons. Such numerical artifacts can be reduced by increasing the mass resolution of the simulations, thereby improving the sampling of the density field on the grid and producing more stable estimates of these quantities.

Our \textit{N}-body simulations were generated using the Adaptive Mesh
Refinement (AMR) code \texttt{ECOSMOG}~\cite{Li:2011vk}, which is itself
based on the AMR code \texttt{RAMSES}~\cite{Teyssier:2001cp}. We used a
simulation box of size $L=1024~\mathrm{Mpc}\,h^{-1}$ with
$N_{p}=1024^{3}$ particles, a domain grid of $1024^{3}$ cells, and a
refinement criterion of $8$. The simulations were initialized at
$z_{\rm ini}=49$ using second-order Lagrangian perturbation theory
(2LPT) initial conditions generated with the code
\texttt{2LPTic}~\cite{Scoccimarro:1997gr,Crocce:2006ve}. To compute the
Lagrangian fields, we generated Gaussian random fields using the
\texttt{FML}
package\footnote{\url{https://github.com/HAWinther/FML/tree/master/FML/RandomFields}},
employing the same GSL random-number generator used by
\texttt{2LPTic}, namely \texttt{gsl\_rng\_ranlxd1}. We ran both
$\Lambda$CDM and F5 simulations using identical cosmological parameters
and matched initial conditions. We further employed the paired-fixed
approach of Ref.~\cite{Angulo:2016hjd} in order to reduce sample
variance. The basis spectra computed from both the \textit{N}-body
simulations and the Zel'dovich realizations were evaluated after
applying a Gaussian smoothing to the initial density field
$\delta_{\rm L}(\boldsymbol{q})$, with smoothing scale
$k_{\rm smooth}=3.14\,h\,\mathrm{Mpc}^{-1}$.
%%%%%%%%%%%%%%%%%%%%%%%%%%%%%%%%%%%%%%%%%
%%%%%%%%%%%%%% FIGURE %%%%%%%%%%%%%%%%%%%
%%%%%%%%%%%%%%%%%%%%%%%%%%%%%%%%%%%%%%%%%
\begin{figure}[h]
    \centering
    \includegraphics[width=\linewidth]{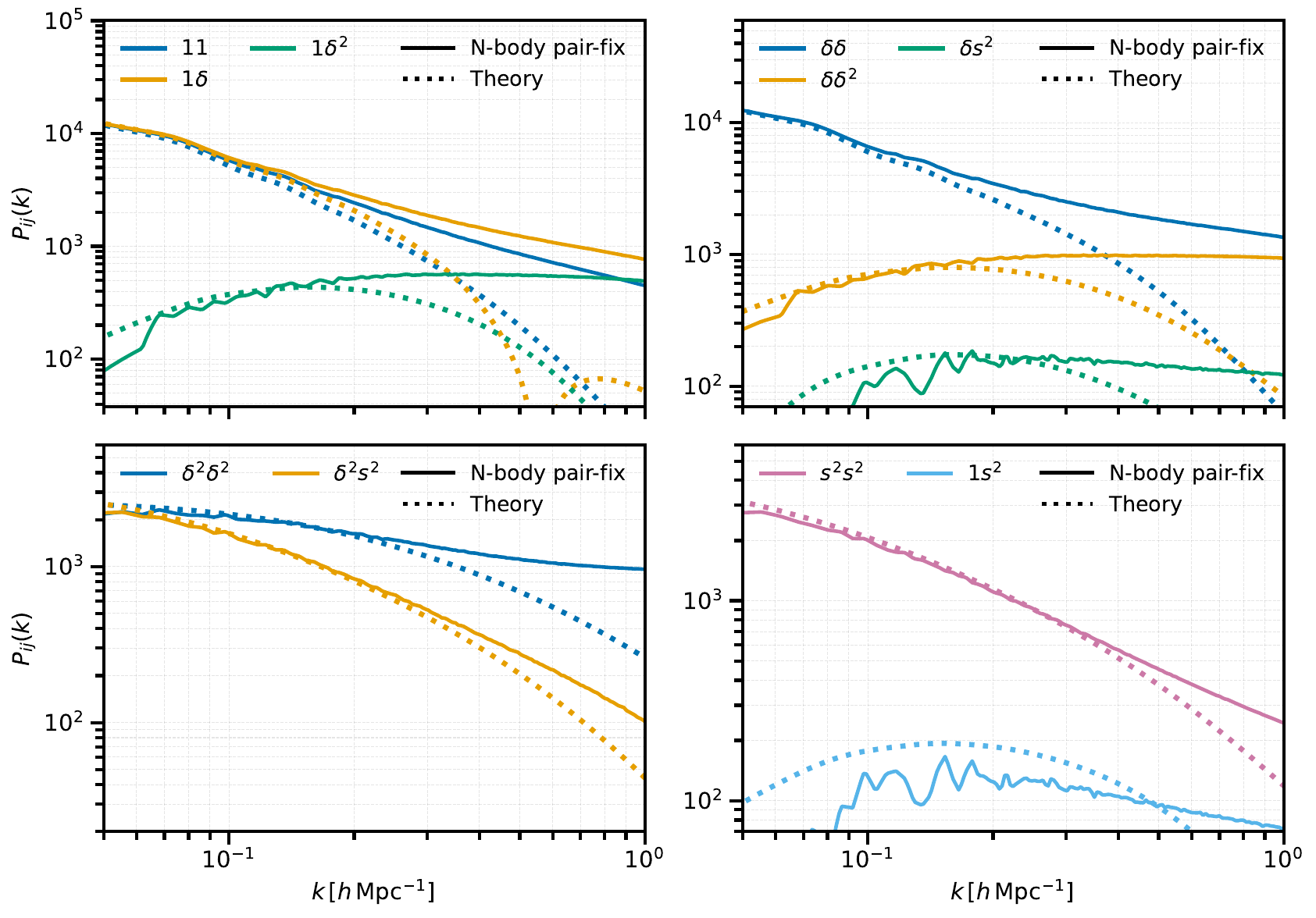}
    \caption{
    Comparison between the HEFT operator spectra measured from pair-fixed \textit{N}-body simulations and the corresponding theoretical predictions obtained from the generalized Lagrangian framework in $f(R)$ gravity. Solid lines denote the spectra measured from the simulations, while dotted lines show the theoretical calculations. The panels display the auto- and cross-spectra associated with the HEFT operators $\delta$, $\delta^{2}$, and $s^{2}$. The pair-fixed simulations substantially suppress cosmic variance on large scales, allowing for a cleaner comparison between theory and measurements.}
    \label{fig:f5_th_nb_P_ij}
\end{figure}
%%%%%%%%%%%%%%%%%%%%%%%%%%%%%%%%%%%%%%%%%
%%%%%%%%%%%%%%%%%%%%%%%%%%%%%%%%%%%%%%%%%
In Fig.~\ref{fig:f5_th_nb_P_ij}, we compare the LPT predictions against
their corresponding \textit{N}-body measurements. The level of residual discrepancies at small wavenumbers is comparable
to that found in previous implementations of the HEFT approach in
$\Lambda$CDM~\cite{Modi:2019qbt,DeRose:2018xdj,Zennaro:2021bwy,
Zhou:2025iiu}.

As can be seen from Figs.~\ref{fig:f5_th_za_P_ij} and \ref{fig:f5_th_nb_P_ij}, some spectra still exhibit noticeable scatter
on large scales. Since we have a direct theoretical implementation for
each LPT basis spectrum, this does not represent a limitation of the
modeling itself, as the large-scale behaviour is analytically controlled
and accurately described. Previous works~\cite{DeRose:2018xdj,
Zhou:2025iiu} have shown that these fluctuations can be substantially
reduced through control-variate techniques~\cite{Chartier:2020pmu,
DeRose:2022zfu}. In this work, however, we will instead adopt a
different strategy. In beyond-$\Lambda$CDM nonlinear modelling of the matter power spectrum, it is customary to emulate the \emph{boost} rather than the power spectrum itself,
\begin{align}
    B(k,z)
    =
    \frac{
    P_{\rm mm}^{\rm NL,\,MG}(k,z|\boldsymbol{\theta})
    }{
    P_{\rm mm}^{\rm NL,\,GR}(k,z|\boldsymbol{\theta})
    },
\end{align}
where $\boldsymbol{\theta}$ represents a set of cosmological and model parameters. This quantity captures the relative modification of the nonlinear matter spectrum with respect to GR. The modified-gravity prediction can then be reconstructed by multiplying the boost by the corresponding GR nonlinear spectrum. Since the GR and MG simulations are generated from matched initial conditions and identical random seeds, this ratio also strongly suppresses cosmic variance. Motivated by this idea, we generalize the same response-ratio strategy to all HEFT operator spectra. For each operator pair $\alpha$, we define the modified-gravity response
\begin{align}
R^\alpha(k,z|\boldsymbol{\theta})
=
\frac{
P_{\rm MG}^{\alpha}(k,z|\boldsymbol{\theta})
}{
P_{\rm GR}^{\alpha}(k,z|\boldsymbol{\theta})
},
\end{align}
this construction substantially reduces the variance of the measured spectra. 

\begin{figure}[h]
    \centering
    \includegraphics[width=\linewidth]{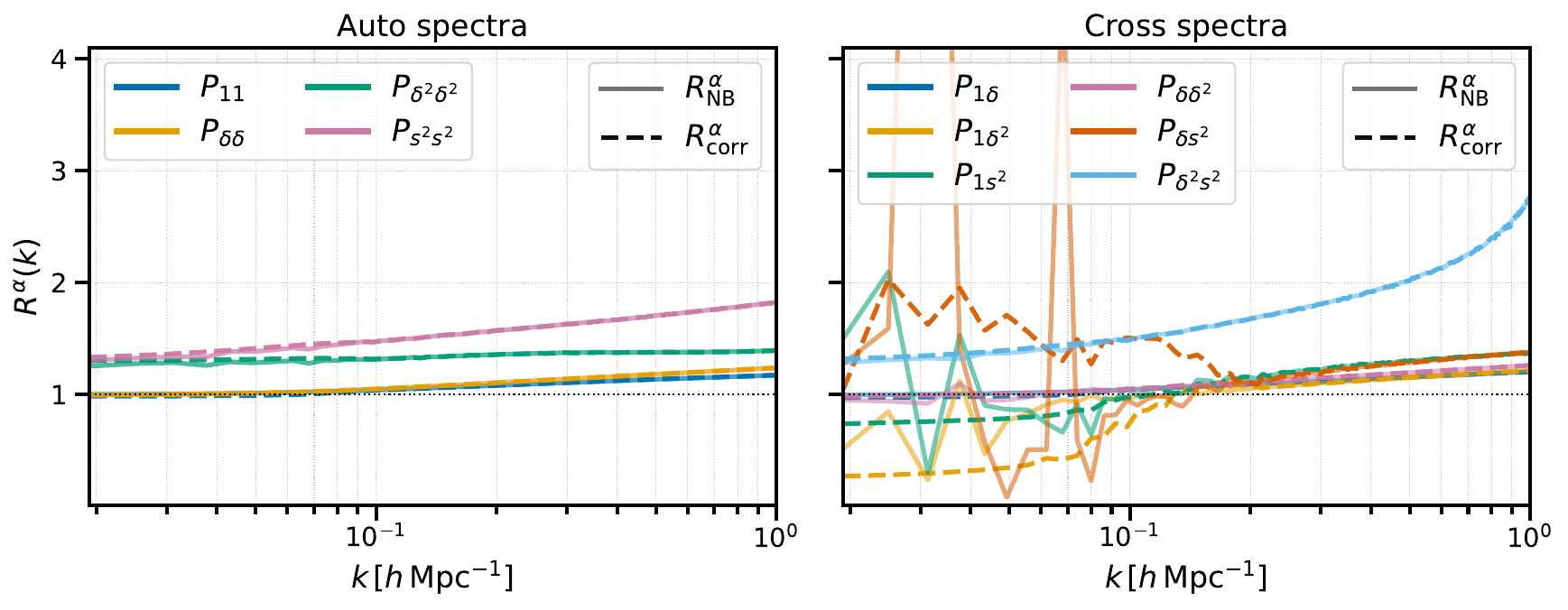}
    \caption{
    Comparison between the direct \textit{N}-body response $R_{\rm NB}^{\alpha}$ and the corrected response $R_{\rm corr}^{\alpha}$ obtained through the stitching and filtering procedure described in the text. The left panel shows the auto spectra, while the right panel displays the cross spectra. The correction procedure significantly reduces the large-scale fluctuations, particularly for the cross spectra, while preserving the nonlinear behaviour at intermediate and mildly nonlinear scales.
    }
    \label{fig:auto_cross_response_comparison}
\end{figure}

The improvement is particularly noticeable for the auto spectra, which remain positive definite and comparatively smooth on large scales. The cross spectra, however, still exhibit residual fluctuations, as they are significantly more sensitive to cancellations and sign changes, see the solid curves in Fig.~\ref{fig:auto_cross_response_comparison}. 

We found that filtering the nonlinear correction relative to the ZA response substantially reduces the noise when compared to filtering the full modified-gravity response directly. 
Our procedure is organized in the following steps. First, we compute the direct \textit{N}-body response from the ratio between the F5 and GR spectra measured in the simulations. Second, we compute the corresponding ZA response using matched Zel'dovich realizations. We then isolate the nonlinear correction beyond the ZA approximation through the ratio
\begin{align}
C^\alpha(k,z|\boldsymbol{\theta})
=
\frac{
R_{\rm NB}^{\alpha}(k,z|\boldsymbol{\theta})
}{
R_{\rm ZA}^{\alpha}(k,z|\boldsymbol{\theta})
}.
\end{align}
This quantity captures the residual nonlinear modification relative to the ZA prediction. In practice, $C^\alpha(k,z|\boldsymbol{\theta})$ is considerably smoother than the full response itself, particularly for the auto spectra, which allows the filtering procedure to be applied more robustly to the correction factor rather than directly to the full modified-gravity response.

For the auto spectra, the correction factor is sufficiently stable. The cross spectra, on the other hand, require an additional low-$k$ smoothing step due to their stronger sensitivity to noise and sign changes. In practice, we fit the large-scale behaviour of the correction factor using a low-order polynomial and smoothly blend this fit into the measured correction over an intermediate transition region. The final filtered correction factor, denoted by $C_{\rm filt}^{\alpha}(k,z|\boldsymbol{\theta})$, is therefore constructed by combining the measured correction with its smooth large-scale fit. This filtering is applied only where necessary: the auto spectra use the direct correction factor, while the cross spectra use the smoothed version at low $k$. The filtered correction is then interpolated onto the Fourier grid used for the HEFT predictions.

Next, we compute the perturbative response using the HEFT basis spectra. Finally, we construct a stitched response that follows the perturbative prediction on very large scales and smoothly transitions to the ZA response towards intermediate and mildly nonlinear scales. Schematically, the stitched response is written as
\begin{align}
R_{\rm stitch}^{\alpha}(k,z|\boldsymbol{\theta})
=
\left[
1-W(k)
\right]
R_{\rm theory}^{\alpha}(k,z|\boldsymbol{\theta})
+
W(k)
R_{\rm ZA}^{\alpha}(k,z|\boldsymbol{\theta}),
\end{align}
where $W(k)$ is a smooth transition function. This construction preserves the reliable large-scale behaviour of perturbation theory while simultaneously taking advantage of the lower-noise ZA response at intermediate scales.
%The agreement is particularly good for the auto spectra, while the cross spectra still retain some residual fluctuations at very low $k$, reflecting their enhanced sensitivity to cancellations and sign changes.}

%%%%%%%%%%%%%%%%%%%%%%%%%%%%%%%%%%%%%%%%%
%%%%%%%%%%%%%% FIGURE %%%%%%%%%%%%%%%%%%%
%%%%%%%%%%%%%%%%%%%%%%%%%%%%%%%%%%%%%%%%%
%%%%%%%%%%%%%%%%%%%%%%%%%%%%%%%%%%%%%%%%%
%%%%%%%%%%%%%%%%%%%%%%%%%%%%%%%%%%%%%%%%%

The final corrected response is reconstructed as
\begin{align}\label{eq:R_corr}
R_{\rm corr}^{\alpha}(k,z|\boldsymbol{\theta})
=
C_{\rm filt}^{\alpha}(k,z|\boldsymbol{\theta})
R_{\rm stitch}^{\alpha}(k,z|\boldsymbol{\theta}) .
\end{align}
In this way, the end response retains the nonlinear information measured from the \textit{N}-body simulations, while its large-scale and mildly nonlinear behaviour are regularized using the smoother perturbative and ZA estimates. Finally, the modified-gravity HEFT spectra are reconstructed by applying the corrected response to the corresponding GR spectra,
\begin{align}
P_{\rm MG}^{\alpha}(k,z|\boldsymbol{\theta})
=
R_{\rm corr}^{\alpha}(k,z|\boldsymbol{\theta})
P_{\rm GR}^{\alpha}(k,z|\boldsymbol{\theta}) .
\end{align}
We illustrate the impact of this procedure in Fig.~\ref{fig:auto_cross_response_comparison}, where we compare the original \textit{N}-body response with the corrected response obtained after applying the filtering and stitching procedure. This procedure provides stable operator-by-operator predictions for the HEFT basis spectra in modified gravity, while simultaneously retaining the variance-reduction benefits of matched GR and MG simulations.

%%%%%%%%%%%%%%%%%%%%%%%%%%%%%%%%%%%%%%%%%
%%%%%%%%%%%%%% FIGURE %%%%%%%%%%%%%%%%%%%
%%%%%%%%%%%%%%%%%%%%%%%%%%%%%%%%%%%%%%%%%
\begin{figure}[h]
    \centering
    \includegraphics[width=\linewidth]{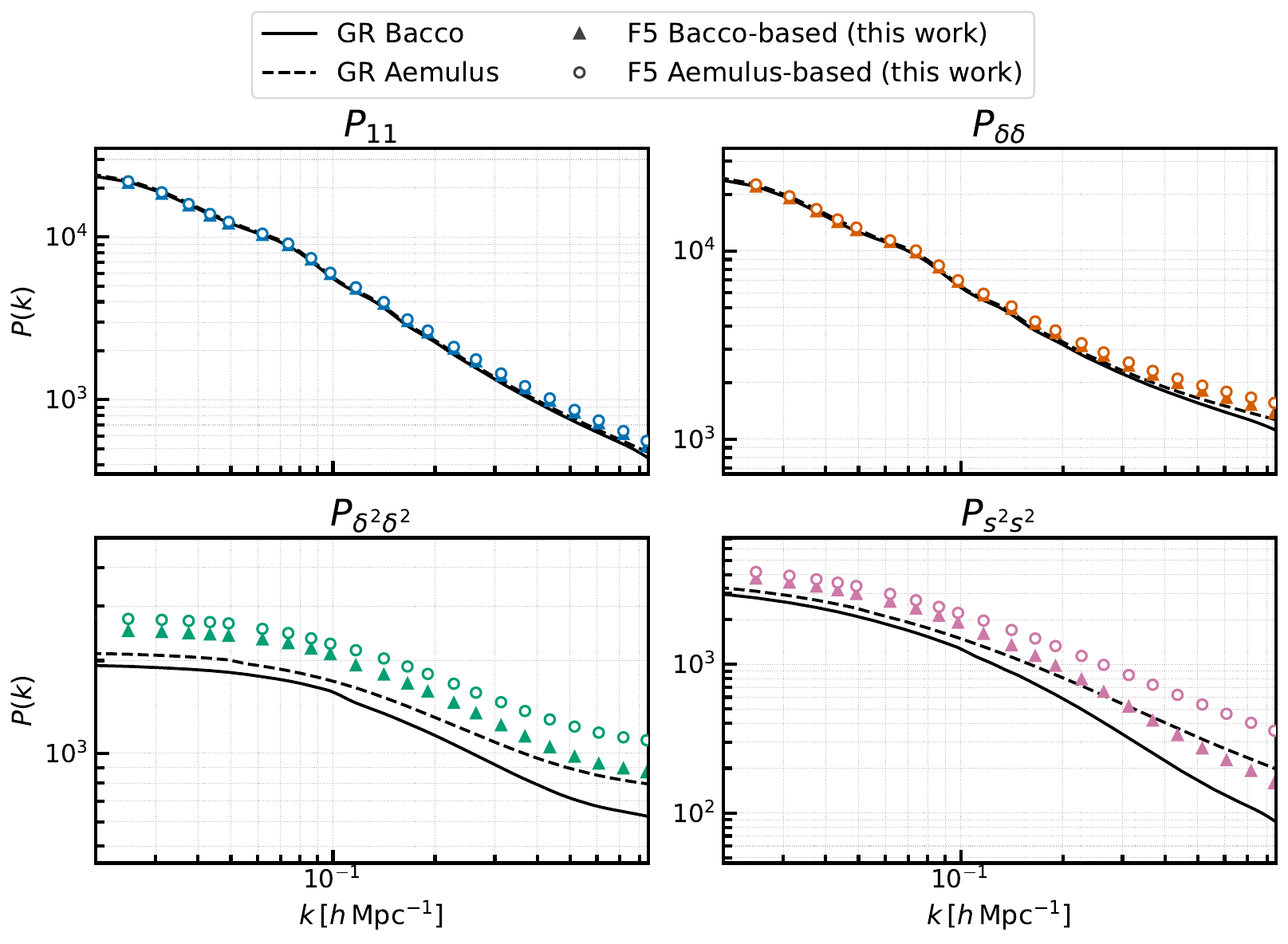}
    \caption{
    Auto-spectrum HEFT basis predictions for F5 gravity obtained using the corrected response procedure of Eq.~(\ref{eq:R_corr}). The modified-gravity spectra are reconstructed from the corresponding GR predictions generated using the \texttt{bacco} and \texttt{Aemulus} emulators.
    }
    \label{fig:auto_F5_pred}
\end{figure}
%%%%%%%%%%%%%%%%%%%%%%%%%%%%%%%%%%%%%%%%%
%%%%%%%%%%%%%%%%%%%%%%%%%%%%%%%%%%%%%%%%%

%%%%%%%%%%%%%%%%%%%%%%%%%%%%%%%%%%%%%%%%%
%%%%%%%%%%%%%% FIGURE %%%%%%%%%%%%%%%%%%%
%%%%%%%%%%%%%%%%%%%%%%%%%%%%%%%%%%%%%%%%%
\begin{figure}[h]
    \centering
    \includegraphics[width=\linewidth]{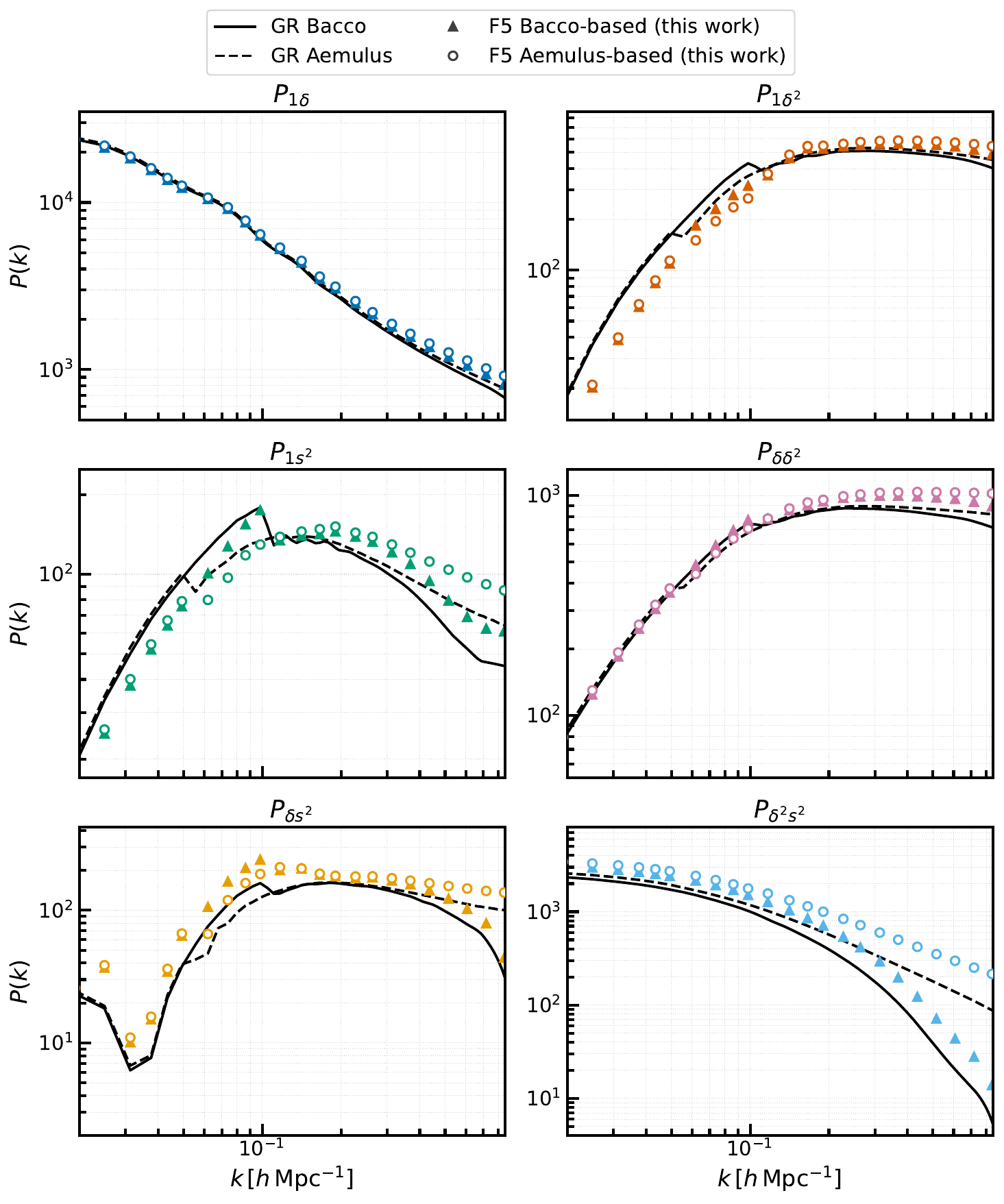}
    \caption{
    Cross-spectrum HEFT basis predictions for F5 gravity reconstructed using the corrected response procedure of Eq.~(\ref{eq:R_corr}). The modified-gravity spectra are obtained by applying the reconstructed response functions to the corresponding GR predictions generated using the \texttt{bacco} and \texttt{Aemulus} emulators.
    }
    \label{fig:cross_F5_pred}
\end{figure}
%%%%%%%%%%%%%%%%%%%%%%%%%%%%%%%%%%%%%%%%%
%%%%%%%%%%%%%%%%%%%%%%%%%%%%%%%%%%%%%%%%%

In Fig.~\ref{fig:auto_F5_pred} we show the auto-spectra, while Fig.~\ref{fig:cross_F5_pred} displays the cross-spectra obtained by applying the reconstructed F5 response functions~(\ref{eq:R_corr}) to the $\Lambda$CDM HEFT basis spectra predicted by the \texttt{aemulus} and \texttt{bacco} emulators, evaluated at the same cosmological parameters used in our F5 simulations. Figures~\ref{fig:auto_F5_pred} and~\ref{fig:cross_F5_pred} illustrate how scale-dependent growth in modified gravity alters the HEFT basis spectra relative to the $\Lambda$CDM case. The modifications observed in $P_{11}$, $P_{\delta\delta}$, and $P_{1\delta}$ are relatively straightforward to interpret, as they exhibit the familiar enhancement of clustering at small scales already well known in the literature. In contrast, spectra involving the operators $\delta_{\rm L}^{2}(\boldsymbol{q})$ and $s^{2}(\boldsymbol{q})$ display a more intricate scale dependence, since these fields are intrinsically sensitive to mode-coupling effects and correlations between long and short wavelength perturbations already at these scales. It is worth noting in the Figures the presence of mild discontinuities visible in some of the spectra. 
These discontinuities originate from the \texttt{bacco} and \texttt{Aemulus} emulators in $\Lambda$CDM. These features are not particularly relevant for practical applications, as they appear at large scales (small wavenumbers), where they remain several orders of magnitude smaller than the dominant HEFT basis contributions that govern the total signal in this regime.

\section{Conclusion}\label{sec:concl}

In this work, we have presented a complete framework for computing both the matter power spectrum and the biased tracer power spectrum within the local Lagrangian bias expansion in modified gravity. We have highlighted the fundamental differences between employing the usual Einstein--de Sitter approximation, commonly adopted in $\Lambda$CDM and mild extensions of it, and performing the full calculation in a generalized cosmology, such as $f(R)$ gravity. The proper derivation of the first-, second-, and symmetrized third-order growth factors allows us to construct the Lagrangian kernels while consistently accounting for the scale- and time-dependent effects characteristic of these theories.

We have also presented, for the first time, the computation of the Hybrid Effective Field Theory basis spectra in $f(R)$ gravity. With this careful derivation of both the analytical predictions and their simulation counterparts, this work lays the foundation for the construction of novel emulators in extensions of $\Lambda$CDM, which we leave for future work. Another important extension of this framework is to investigate how the bias parameters in generalized cosmologies correlate with galaxy formation models~\cite{Kokron:2021faa,Zennaro:2021pbe}. For this purpose, a careful analysis of hydrodynamical simulations, such as those in~\cite{Arnold:2019vpg}, will be of great interest.

Furthermore, in this work we have focused exclusively on real-space spectra. However, galaxy observations are fundamentally performed in redshift space, making an accurate extension of this modelling essential for future analyses. Fortunately, the LPT formalism developed here provides a natural and straightforward connection to redshift-space modelling, as shown in~\cite{Matsu2}.

\acknowledgments

This work used the DiRAC@Durham facility managed by the Institute for
Computational Cosmology on behalf of the STFC DiRAC HPC Facility
(\url{https://www.dirac.ac.uk}). The equipment was funded by BEIS
capital funding via STFC capital grants ST/K00042X/1, ST/P002293/1,
ST/R002371/1, and ST/S002502/1, Durham University, and STFC operations
grant ST/R000832/1. DiRAC is part of the UK National e-Infrastructure. This work also made use of the CHE cluster, managed and funded by COSMO/CBPF/MCTI, with financial support from CNPq, FINEP and FAPERJ. BL is supported by the ERC Advanced Grant `UNCA' (UKRI Frontiers Research Guarantee No.~EP/Z533877/1) and the UK STFC Consolidated Grant ST/X001075/1. KK is  supported by STFC grant number ST/B001175/1. 

For the purpose of open access, we have applied a Creative Commons Attribution (CC BY) licence to any Author Accepted
Manuscript version arising. Supporting research data are available on reasonable request from the authors.

\paragraph{Data availability} Supporting research data are available on reasonable request from the corresponding author.

%%%%%%%%%%%%%%
\appendix 
\section{General Fast Schmitfull Integrals}\label{app:fast_integrals_mg}

In this Appendix we present a way to apply the same strategy devised in~\cite{Schmittfull:2016jsw} but for the case of kernels that can be expresend in analytical terms. We begin by expanding the full three-dimensional integrand:
\begin{equation}
    Q(k)=\int \frac{d^{3}p}{(2\pi)^{3}}\mathcal{V}\left(p,|\boldsymbol{k-p}|, \boldsymbol{\hat{p}}\cdot \widehat{\boldsymbol{k - p}}\right)P_{\rm L}(p)P_{\rm L}(|\boldsymbol{k-p}|),
\end{equation}
where $\mathcal{V}\left(p,|\boldsymbol{k-p}|, \boldsymbol{\hat{p}}\cdot \widehat{\boldsymbol{k - p}}\right)$ is the corresponding kernel appearing in $Q_{n}$ or $R_{n}$ for a general cosmology. This function can be expressed as an infinite sum of Legendre polynomials:
\begin{equation}
    \mathcal{V}\left(p,|\boldsymbol{k-p}|, \mu\right)
    =
    \sum_{\ell=0}^{\infty} V_{\ell}(k,p)\,P_{\ell}(\mu),
\end{equation}
where $\mu = \boldsymbol{\hat{p}}\cdot \widehat{\boldsymbol{k - p}}$, and the coefficients are given by the projection onto the Legendre basis:
\begin{equation}
    V_{\ell}(k,p)
    =
    \frac{2\ell + 1}{2}
    \int_{-1}^{1}d\mu \;
    \mathcal{V}\!\left(p,\sqrt{k^{2}+p^{2}-2kp\,\mu}, \mu\right)
    P_{\ell}(\mu).
\end{equation}
Substituting this expansion into the full integral, we obtain
\begin{align}\label{eq:general_r_q_xilm}
    Q(k)
    &=
    \sum_{\ell=0}^{\infty}
    \int \frac{d^{3}p}{(2\pi)^{3}}
    V_{\ell}(k,p)\,P_{\ell}(\mu)
    P_{\rm L}(p)P_{\rm L}(|\boldsymbol{k-p}|).
\end{align}
Therefore, even though a finite Legendre decomposition is not available, the integrand can still be expanded as an infinite series of Legendre polynomials. 

To proceed further, one must additionally represent each multipole coefficient $V_\ell(k,p)$ in a separable radial form. In full generality, this can be written as
\begin{equation}
    V_{\ell}(k,p)
    \approx
    \sum_{\alpha} \lambda_{\alpha}^{(\ell)}(k)\, f_{\alpha}(p),
\end{equation}
or more generally, at the level of the full kernel,
\begin{equation}
    \mathcal{V}(p,|\boldsymbol{k-p}|,\mu)
    \approx
    \sum_{\ell=0}^{\infty}\sum_{\alpha}
    \lambda_{\alpha}^{(\ell)}(k)\,
    f_{\alpha}(p)\,g_{\alpha}(|\boldsymbol{k-p}|)\,
    P_{\ell}(\mu),
\end{equation}
where $\{f_\alpha,g_\alpha\}$ define a chosen separable radial basis, and the coefficients $\lambda_{\alpha}^{(\ell)}(k)$ are obtained by numerical projection or interpolation.

Substituting this expansion into Eq.~\eqref{eq:general_r_q_xilm} gives
\begin{equation}
    Q(k)
    =
    \sum_{\ell=0}^{\infty}\sum_{\alpha} \lambda_{\alpha}^{(\ell)}(k)
    \int \frac{d^{3}p}{(2\pi)^{3}}
    f_{\alpha}(p)\,g_{\alpha}(|\boldsymbol{k-p}|)\,P_{\ell}(\mu)
    P_{\rm L}(p)P_{\rm L}(|\boldsymbol{k-p}|).
\end{equation}

Each term in the sum can then be written in configuration space by defining generalized correlation functions associated with the basis functions $f_\alpha$ and $g_\alpha$:
\begin{equation}
    \Xi_{f_\alpha}^{\,\ell}(r)
    =
    i^\ell \int \frac{d^{3}p}{(2\pi)^{3}}
    e^{-i\boldsymbol{p}\cdot\boldsymbol{r}}
    f_{\alpha}(p)\,
    P_{\ell}(\boldsymbol{\hat{p}}\cdot \boldsymbol{\hat{r}})
    P_{\rm L}(p),
\end{equation}
and analogously for $\Xi_{g_\alpha}^{\,\ell}(r)$. One then obtains
\begin{equation}
    Q(k)
    =
    \sum_{\ell=0}^{\infty}\sum_{\alpha}
    \lambda_{\alpha}^{(\ell)}(k)
    \frac{(-1)^{\ell}4\pi}{(2\pi)^{3}}
    \int dr\,r^{2}j_{0}(kr)\,
    \Xi_{f_\alpha}^{\,\ell}(r)\,
    \Xi_{g_\alpha}^{\,\ell}(r).
\end{equation}

A particularly convenient and widely used choice of separable basis is provided by the FFTLog decomposition, in which the radial dependence is expanded in complex power laws,
\begin{equation}
    f_{\alpha}(p) = p^{\nu_\alpha}, \qquad
    g_{\alpha}(p) = p^{\nu_\alpha}, \qquad \nu_\alpha \in \mathbb{C}.
\end{equation}
In this case, the generalized correlation functions reduce to
\begin{equation}
    \xi_{\nu}^{\ell}(r)
    =
    \int \frac{dk}{2\pi^{2}} k^{\nu+2} j_{\ell}(kr) P_{\rm L}(k),
\end{equation}
that correspond to Hankel transforms of order $\ell$. The final expression becomes
\begin{equation}\label{eq:Q_gen}
    Q(k)
    =
    \sum_{\ell=0}^{\infty}\sum_{\alpha,\beta}
    c_{\alpha\beta}^{(\ell)}(k)
    \frac{(-1)^{\ell}4\pi}{(2\pi)^{3}}
    \int dr\,r^{2}j_{0}(kr)\,
    \xi_{\nu_\alpha}^{\ell}(r)\,
    \xi_{\nu_\beta}^{\ell}(r).
\end{equation}

The generalized correlation functions arise once each multipole component of the kernel is decomposed into separable radial factors in $p$ and $|\boldsymbol{k-p}|$. The standard $\xi_n^\ell(r)$ representation corresponds to the particular choice $f_\alpha(p)=p^{n_1}$ and $g_\alpha(p)=p^{n_2}$, while an FFTLog expansion provides a natural and efficient implementation in terms of Hankel transforms.

\bibliographystyle{JHEP}
\bibliography{bib_my}

\end{document}